\documentclass[letter]{aa}  

\usepackage{graphicx}
\usepackage{txfonts}
\usepackage{graphicx}   
\usepackage{amsmath}    
\usepackage{wasysym}
\usepackage{placeins}

\begin{document} 
\newcommand{\irasstar}{IRAS 04125+2902}
\DeclareRobustCommand{\orcidicon}{%
    \begin{tikzpicture}
    \draw[lime, fill=lime] (0,0) 
    circle [radius=0.16] 
    node[white] {{\fontfamily{qag}\selectfont \tiny ID}};
    \draw[white, fill=white] (-0.0625,0.095) 
    circle [radius=0.007];
    \end{tikzpicture}
    \hspace{-2mm}
}

\newcommand*{\tabhead}[1]{\multicolumn{1}{c}{\bfseries #1}}

\newcommand{\mystar}{{\Large {\fontfamily{lmr}\selectfont$\star$}}}
\newcommand{\pds}{PDS~70}

\newcommand{\jgrp}{JGR-Planets}
\newcommand{\rsos}{Royal Soc. Open Sci.}
\newcommand{\astrolett}{Astron. Lett.}
\newcommand{\psj}{Planet. Sci. J.}
\newcommand{\jatis}{J. Astron. Telesc. Instrum. Syst.}
\newcommand{\mgtwo}{Mg\,II}
\newcommand{\ctwo}{C\,II}
\newcommand{\molh}{H$_2$}
\newcommand{\lyalph}{Ly-$\alpha$}
\newcommand{\fetwo}{Fe\,II}

\newcommand{\lognh}{$\log(N_{\rm H~I})$}
\newcommand{\oi}{[O\,I]}
\newcommand{\suii}{[S\,II]}
\newcommand{\nii}{[N\,II]}
\newcommand{\nai}{Na\,I}
\newcommand{\caii}{Ca\,II}
\newcommand{\mdotacc}{$\cdot{M}_{\rm acc}$}
\newcommand{\mdotwind}{$\cdot{M}_{\rm wind}$}
\newcommand{\hei}{He\,I}
\newcommand{\lyalpha}{Ly-$\alpha$}
\newcommand{\halpha}{H$\alpha$}
\newcommand{\mgi}{Mg\,I}
\newcommand{\sii}{Si\,I}
\newcommand{\water}{H$_2$O}
\newcommand{\methane}{CH$_4$}
\newcommand{\cotwo}{CO$_2$}
\newcommand{\htwo}{H$_2$}


\newcommand{\rosat}{\emph{Rosat}}
\newcommand{\galex}{\emph{GALEX}}
\newcommand{\tess}{\emph{TESS}}
\newcommand{\plato}{\emph{PLATO}}
\newcommand{\gaia}{\emph{Gaia}}
\newcommand{\ktwo}{\emph{K2}}
\newcommand{\jwst}{\emph{JWST}}
\newcommand{\kepler}{\emph{Kepler}}
\newcommand{\corot}{\emph{CoRoT}}
\newcommand{\hipp}{\emph{Hipparcos}}
\newcommand{\spitzer}{\emph{Spizter}}
\newcommand{\herschel}{\emph{Herschel}}
\newcommand{\hst}{\emph{HST}}
\newcommand{\wise}{\emph{WISE}}
\newcommand{\swift}{\emph{Swift}}
\newcommand{\chandra}{\emph{Chandra}}
\newcommand{\xmm}{\emph{XMM-Newton}}

\newcommand{\twa}{TW Hydra}
\newcommand{\bpic}{$\beta$~Pictoris}
\newcommand{\abdor}{AB~Doradus}
\newcommand{\rup}{Ruprecht~147}
\newcommand{\etacha}{$\eta$\,Chamaeleontis}
\newcommand{\usco}{Upper~Sco}
\newcommand{\rhooph}{$\rho$~Oph}

\newcommand{\doar}{DoAr\,25}
\newcommand{\epcha}{EP\,Cha}
\newcommand{\rylup}{RY\,Lup}
\newcommand{\hdtwofour}{HD\,240779}

\newcommand{\sigrv}{$\sigma_{\rm RV}$}

\newcommand{\msunyr}{$\rm{M_{\sun} \, yr^{-1}}$}
\newcommand{\etal}{\mbox{\rm et al.~}}
\newcommand{\ms}{\mbox{m\,s$^{-1}~$}}
\newcommand{\kms}{\mbox{km\,s$^{-1}~$}}
\newcommand{\ks}{\mbox{km\,s$^{-1}~$}}
\newcommand{\kse}{\mbox{km\,s$^{-1}$}}
\newcommand{\mse}{\mbox{m\,s$^{-1}$}}
\newcommand{\msy}{\mbox{m\,s$^{-1}$\,yr$^{-1}~$}}
\newcommand{\msye}{\mbox{m\,s$^{-1}$\,yr$^{-1}$}}
\newcommand{\msun}{M$_{\odot}$}
\newcommand{\msune}{M$_{\odot}$}
\newcommand{\rsun}{R$_{\odot}$}
\newcommand{\lsun}{L$_{\odot}~$}
\newcommand{\rsune}{R$_{\odot}$}
\newcommand{\mjup}{M$_{\rm JUP}~$}
\newcommand{\mjupe}{M$_{\rm JUP}$}
\newcommand{\msat}{M$_{\rm SAT}~$}
\newcommand{\msate}{M$_{\rm SAT}$}
\newcommand{\mnep}{M$_{\rm NEP}~$}
\newcommand{\mnepe}{M$_{\rm NEP}$}
\newcommand{\mearth}{$M_{\oplus}$}
\newcommand{\mearthe}{$M_{\oplus}$}
\newcommand{\rearth}{$R_{\oplus}$}
\newcommand{\rearthe}{$R_{\oplus}$}
\newcommand{\rjup}{R$_{\rm JUP}~$}
\newcommand{\msinie}{$M_{\rm p} \sin i$}
\newcommand{\vsinie}{$V \sin i$}
\newcommand{\mbsini}{$M_b \sin i~$}
\newcommand{\mcsini}{$M_c \sin i~$}
\newcommand{\mdsini}{$M_d \sin i~$}
\newcommand{\chisq}{$\chi_{\nu}^2$}
\newcommand{\chinu}{$\chi_{\nu}$}
\newcommand{\chinusq}{$\chi_{\nu}^2$}
\newcommand{\arel}{$a_{\rm rel}$}
\newcommand{\feh}{\ensuremath{[\mbox{Fe}/\mbox{H}]}}
\newcommand{\rphk}{\ensuremath{R'_{\mbox{\scriptsize HK}}}}
\newcommand{\lrphk}{\ensuremath{\log{\rphk}}}
\newcommand{\cs}{$\sqrt{\chi^2_{\nu}}$}
\newcommand{\etaearth}{$\mathbf \eta_{\oplus} ~$}
\newcommand{\etaearthe}{$\mathbf \eta_{\oplus}$}
\newcommand{\searth}{$S_{\bigoplus}$}
\newcommand{\mdotyr}{$M_{\odot}$~yr$^{-1}$}
\newcommand{\micron}{$\mu$m}

\newcommand{\sini}{\ensuremath{\sin i}}
\newcommand{\msini}{\ensuremath{M_{\rm p} \sin i}}
\newcommand{\mplsini}{\ensuremath{\mpl\sin i}}
\newcommand{\teff}{\ensuremath{T_{\rm eff}}}
\newcommand{\teq}{\ensuremath{T_{\rm eq}}}
\newcommand{\logg}{\ensuremath{\log{g}}}
\newcommand{\vsini}{\ensuremath{v \sin{i}}}
\newcommand{\ebv}{E($B$-$V$)}

\newcommand{\Kepler}{\emph{Kepler}~}
\newcommand{\Keplere}{\emph{Kepler}}
\newcommand{\blender}{{\tt BLENDER}~}

\newcommand{\kp}{\ensuremath{\mathrm{Kp}}}
\newcommand{\rstar}{\ensuremath{R_\star}}
\newcommand{\mstar}{\ensuremath{M_\star}}
\newcommand{\loggstar}{\ensuremath{\logg_\star}}
\newcommand{\lstar}{\ensuremath{L_\star}}
\newcommand{\astar}{\ensuremath{a_\star}}
\newcommand{\loglstar}{\ensuremath{\log{L_\star}}}
\newcommand{\rhostar}{\ensuremath{\rho_\star}}

\newcommand{\rp}{\ensuremath{R_{\rm p}}}
\newcommand{\rpl}{\ensuremath{r_{\rm P}}~}
\newcommand{\rple}{\ensuremath{r_{\rm P}}}
\newcommand{\mpl}{\ensuremath{m_{\rm P}}~}
\newcommand{\lpl}{\ensuremath{L_{\rm P}}~}
\newcommand{\rhopl}{\ensuremath{\rho_{\rm P}}~}
\newcommand{\loggpl}{\ensuremath{\logg_{\rm P}}~}
\newcommand{\logrpl}{\ensuremath{\log r_{\rm P}}~}
\newcommand{\logrple}{\ensuremath{\log r_{\rm P}}}
\newcommand{\loga}{\ensuremath{\log a}~}
\newcommand{\logmpl}{\ensuremath{\log m_{\rm P}}~}

\newcommand{\fludensunits}{ergs s$^{-1}$ cm$^{-2}$ \AA$^{-1}$}

\newcommand{\rr}{$\mathcal{R}$}

\def\Bl{\hbox{$B_{\rm \ell}$}}
\def\Porb{\hbox{$P_{\rm orb}$}}
\def\Prot{\hbox{$P_{\rm rot}$}}
\def\mic{\hbox{$\mu$m}}
\def\chisqr{\hbox{$\chi^2_{\rm r}$}}
\def\vsini{\hbox{$v \sin i$}}
\def\kms{\hbox{km\,s$^{-1}$}}
\def\teff{\hbox{$T_{\rm eff}$}}
\def\logg{\hbox{$\log g$}}
\def\ms{\hbox{m\,s$^{-1}$}}
\def\mjup{\hbox{${\rm M}_{\jupiter}$}}
\def\gpcc{\hbox{g\,cm$^{-3}$}}
\def\rcor{\hbox{$r_{\rm cor}$}}
\def\rmag{\hbox{$r_{\rm mag}$}}
\def\rstar{\hbox{$R_{\star}$}}
\def\rsun{\hbox{${\rm R}_{\odot}$}}
\def\mspy{\hbox{${\rm M}_{\odot}$\,yr$^{-1}$}}
\def\lspy{\hbox{${\rm L}_{\odot}$\,yr$^{-1}$}}
\def\emr{}

\title{Mass, Gas, and Gauss around a T Tauri Star with SPIRou}

\author{
J.-F.~Donati\inst{\ref{irap}\thanks{jean-francois.donati@irap.omp.eu}}
\and
E.~Gaidos\inst{\ref{hawaii},\ref{vienna}}
\and
C. Moutou\inst{\ref{irap}} 
\and 
P. I. Cristofari\inst{\ref{cfa}} 
\and 
L. Arnold\inst{\ref{cfht}}
\and
M. G. Barber\inst{\ref{unc}}
\and
A. W. Mann\inst{\ref{unc}}
}
   
\institute{
Universit\'e de Toulouse, CNRS, IRAP, 14 avenue Belin, 31400 Toulouse, France\label{irap}
\and
Department of Earth Sciences, University of Hawai'i at M\={a}noa, Honolulu, Hawai'i 96822 USA\label{hawaii}
\and
Institute for Astrophysics, University of Vienna, 1180 Vienna, Austria\label{vienna}
\and
Center for Astrophysics | Harvard \& Smithsonian, 60 Garden street, Cambridge, MA 02138, United States\label{cfa}
\and
Canada-France-Hawaii Telescope, 65-1238 Mamalahoa Hwy, Kamuela, HI 96743, USA\label{cfht}
\and
Department of Physics and Astronomy, University of North Carolina Chapel Hill, Chapel Hill, NC 27599 USA\label{unc}
}
\date{Received 2025 March 18; accepted 2025 May 15}
\abstract 
{Studies of young planets help us understand planet evolution and investigate important evolutionary processes such as atmospheric escape.  We monitored \irasstar, a 3~Myr-old T~Tauri star with a transiting planet and a transitional disk, with the SPIRou infrared spectropolarimeter on the Canada-France-Hawaii Telescope.  Using these data, we constrained the mass and density of the Jupiter-size companion to $<$0.16~\mjup\ and $<$0.23~\gpcc, respectively (90\% upper limits).  These rule out a Jovian-like object and support the hypothesis that it is an ancestor to the numerous sub-Neptunes found around mature stars.   We unambiguously detect magnetic fields at the stellar surface, small-scale fields reaching 1.5~kG and the large-scale field mostly consisting of a 0.80$-$0.95~kG dipole inclined by 5$-$15\degr\ to the rotation axis.
Accretion onto the star is low and/or episodic at a maximum rate of $\simeq$$10^{-11}$~\mspy, indicating that \irasstar\ is most likely in a magnetic ``propeller" regime, possibly maintaining the star's slow rotation (11.3~d).   We discover persistent Doppler-shifted absorption in a metastable \hei\ line, clear evidence for a magnetized wind from a gaseous inner disk.  Variability in absorption suggests structure in the disk wind that could reflect disk-planet interactions.}

\maketitle
%

\section{Introduction}

Observing young planets and comparing them to their older counterparts is a powerful method to study planet evolution {\emr \citep{Dai2024,Barber2024b,Fernandes2025}}.  Also, some evolutionary processes such as atmospheric escape can be studied directly by observations of young planets where rates are expected to be fastest.  Most exoplanets on close-in orbits have been discovered by the photometric transit method, and many of these have been monitored with Doppler radial velocity (RV) spectrometers to confirm their existence and measure their masses.  But very young stars are rapidly rotating, highly magnetically active and can still be accreting from primordial disks.  As a result, they are photometrically and spectroscopically variable, inhibiting detection by the transit and RV methods.  Only a handful of young planets have been found by either method in nearby $<$20~Myr star-forming regions, and thus we know little about the progenitors of planets found around mature stars.

A transiting, Jupiter-size planet on a 8.8-d (0.08~au) orbit was recently discovered in \tess\ photometry of \irasstar, a 0.7\,\msun\ member of the Taurus star-forming region \citep{Barber2024}.   At $3.0\pm0.4$~Myr \citep{Luhman2025}, the host star is $3\times$ younger than the previous record holder.   Sparse RV observations limit the planet's mass to no more than that of Jupiter \citep{Barber2024}.  Rather than an actual Jovian planet it could be a highly inflated ancestor to (sub)-Neptunes orbiting older stars, but a more stringent mass constraint is needed.  

The star hosts a transitional disk responsible for excess emission at $\lambda>8$~\micron\ \citep{Luhman2009,Furlan2009,Espaillat2015}.  Paradoxically, mm-wave imaging reveals a nearly face-on ($i\sim30^{\circ}$) disk with a 20~au cavity, extending to 50--60~au \citep{Espaillat2015}, and highly inclined to both the orbit of the planet and of its distant stellar companion \citep{Barber2024}.  $U$-band emission ambiguously suggests accretion \citep{Espaillat2015}, and extinction along the line of sight (LOS) is $A_V = 2.7\pm0.2$~mag \citep{Espaillat2015,Barber2024}, larger than that from dust maps \citep[1.85,][]{Green2018}.  Both hint at an inner disk but more direct evidence is required.

Finally, \irasstar\ is slowly rotating (11.3~d) for a Taurus star \citep{Rebull2020}.  This could reflect efficient magnetic braking by the disk \citep[e.g.,][]{Gehrig2023}, but a measurement of the stellar field is required to evaluate this.  Otherwise, its slow rotation could be somehow connected to the existence/detection of the planet or stellar companion.  Here, we use spectropolarimetric monitoring of \irasstar\ to detect and model the stellar magnetic field, better constrain the planet mass, and unambiguously detect an inner gaseous disk.

\section{Observations and data reduction}
\label{sec:observations}

We observed \irasstar\ with SPIRou \citep{Donati2020}, the near-infrared precision velocimeter / spectropolarimeter on the 3.6-m Canada-France-Hawaii Telescope (CFHT) atop Maunakea in Hawaii.  SPIRou collects unpolarized and polarized stellar spectra over $\lambda$=0.95--2.50~\mic\ at a resolving power of 70\,000 in a single exposure.  We obtained unpolarized (Stokes $I$) and circularly polarized (Stokes $V$) spectra of \irasstar\ at 42 epochs over 118~d between 2024 Oct~24 to 2025 Feb~19 (RUNIDs \#24BH02 and \#25AD98, PI E.~Gaidos), with  observations on Nov~15 and Nov~24 obtained in poor weather discarded.   Seventeen observations were obtained during \tess\ monitoring of the star in Sector 86 (2024 Nov~21 to Dec~18, Fig.~\ref{fig:tess}).  Each SPIRou observation consists of sequences of 4 sub-exposures obtained at different azimuths of the polarimeter Fresnel rhomb retarders.  This procedure can remove systematics in polarization spectra \citep[to first order,][]{Donati1997}. Each recorded sequence yields a pair of Stokes $I$ and $V$ spectra, plus a null polarization diagnostic of potential instrument or reduction issues.

{\emr All spectra of \irasstar\ were processed with the SPIRou reduction package \texttt{APERO} \citep{Cook2022}. 
We applied Least-Squares Deconvolution \citep[LSD,][]{Donati1997} to all reduced spectra} 
with a line mask computed with the VALD-3 database \citep{Ryabchikova2015} for a set of atmospheric parameters ($\teff=4000$~K and $\logg=4.0$) matching those of \irasstar, selecting only atomic lines deeper than 10\% of the continuum level, for a total of $\simeq$1500 lines (average wavelength and Land\'e factor of 1750~nm and 1.2).  We also constructed a second LSD mask with only the $\simeq$300 CO lines of the CO bandhead, known to be insensitive to magnetic fields and thus less affected by stellar activity.  We finally applied the Line-By-Line analysis \citep[LBL,][]{Artigau2022} to derive RVs from all (i.e., $\simeq$25000) spectral features simultaneously (using the median spectrum of Gl~846 as reference), and to quantify the small temperature variations $dT$ that result from starspots \citep{Artigau2024}.  The observation log is given in Table\ref{tab:log}. 

\begin{figure*}[h!]
    \centering
    \includegraphics[width=0.85\textwidth]{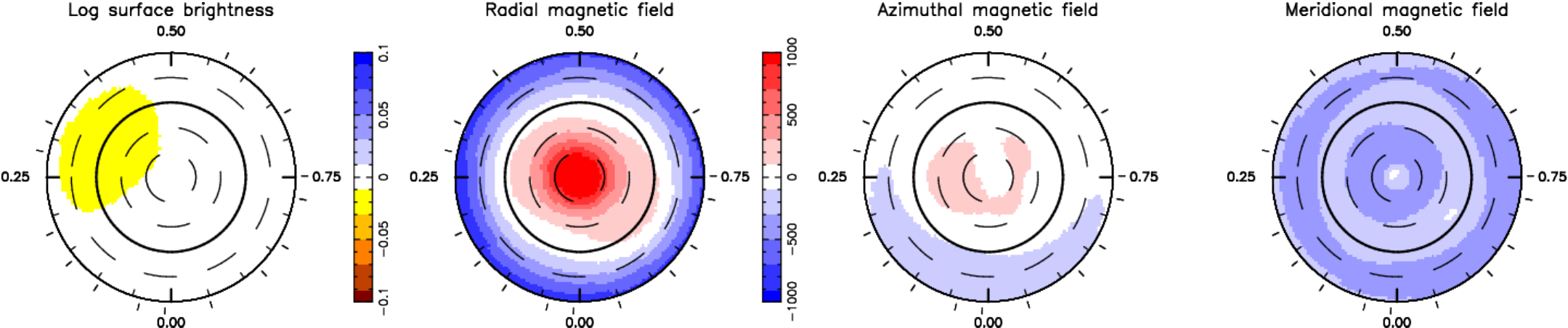}
    \caption{Maps of the reconstructed brightness and large-scale magnetic field of \irasstar\ derived with ZDI from 16 Stokes $I$ and $V$ LSD profiles and $dT$ measurements collected between BJD 2460648 and 2460668. Left to right: logarithmic brightness relative to the quiet photosphere, and the radial, azimuthal and meridional field components in Gauss and spherical coordinates.  Maps are shown in a flattened polar projection down to latitude $-60$\degr, with the north pole at the center and the equator depicted as a bold line.  Outer ticks indicate phases of observations.      }
    \label{fig:mapB}
\end{figure*}

\section{Characterizing the magnetic field } 
\label{sec:zdi}

For each pair of Stokes $I$ and $V$ LSD profiles, we computed the longitudinal magnetic field \Bl, i.e., the LOS-projected component of the vector magnetic field averaged over the visible hemisphere, following \citet{Donati1997}.  This yielded \Bl\ values from $-$47 to 101~G (median 31~G) with uncertainties from 9 to 18~G (median 10~G), confirming that the field is clearly detected at the stellar surface.  Fitting these values by Gaussian Process regression (GPR) with a quasi-periodic kernel \citep[][see  Sec.~\ref{sec:appB} for more details]{Haywood2014,Rajpaul2015} and a Monte-Carlo Markov Chain (MCMC) process, we re-estimated the rotation period $\Prot=11.35\pm0.04$~d as well as the coherence timescale of \Bl\ ($97\pm30$~d), with all inferred hyper-parameters listed in Table~\ref{tab:gpr}.  Figure~\ref{fig:bell} clearly shows \Bl\ evolving over the course of our observations.  

We proceeded in the same way with $dT$ (see Fig.~\ref{fig:bell}),  
showing clear quasi-periodic variation from $-$63 to 29~K with uncertainties between  1.3 and 2.4~K (median 1.4~K), and yielding a slightly shorter $\Prot=11.28\pm0.05$~d and an identical coherence timescale.  As expected, the minimum temperature occurs in conjunction with the minimum brightness in the contemporaneous TESS light curve, at, e.g., BJD = 2460645.8 and 2460657.1.  

We also analyzed the median intensity spectrum of \irasstar\ with {\tt ZeeTurbo} \citep{Cristofari2023} to reassess the main stellar parameters and estimate the intensity of its small-scale magnetic field.  Assuming solar abundances, we derive a photospheric temperature of $\teff=3889\pm30$~K and a logarithmic surface gravity of $\logg=3.92\pm0.05$, in good agreement with previous estimates \citep{Lopez-Valdivia2021,Barber2024}.  In addition, we find that the small-scale field reaches $<B>=1.52\pm0.05$~kG, slightly larger but still consistent with the previous estimate of $1.36\pm0.32$~kG by \citet{Lopez-Valdivia2021}.  More specifically, we find that the median spectrum is best reproduced by a linear combination of synthetic spectra associated with fields of 0, 2 and 4~kG and respective filling factors of $0.27\pm0.02$, $0.70\pm0.03$ and $0.03\pm0.01$.  

We modeled the Stokes $I$ and $V$ LSD profiles of \irasstar\ using Zeeman-Doppler imaging (ZDI) to simultaneously reconstruct the topology of the large-scale magnetic field and the distribution of brightness features at the stellar surface.  We also used our $dT$ measurements to infer a relative light curve at SPIRou wavelengths (using the Planck function to convert temperature changes into brightness fluctuations) which we adjusted at the same time as the Stokes $I$ and $V$ LSD profiles.  Given the relatively fast evolution of \Bl\ (see Fig.~\ref{fig:bell}), we split our data into three subsets covering intervals 2024 Oct-Nov, 2024 Dec and 2025 Feb, each spanning only a few rotation periods.  We used the same code as in previous ZDI analyses of SPIRou data \citep[e.g.,][]{Donati2023,Donati2024} to reconstruct the brightness distribution and the large-scale magnetic field at the rotating stellar surface from phase-resolved sets of Stokes $I$ and $V$ LSD profiles and $dT$ photometry (Sec.~\ref{sec:appC}).  

We assumed \irasstar\ rotates as a solid body with $\Prot=11.35$~d, and set the inclination angle of the rotation axis to the LOS to $i=80\degr$ (to limit mirroring between hemispheres in the ZDI inversion process) and thus the LOS projected equatorial rotation velocity \vsini\ to 6.3~\kms\ \citep[the most likely value given \Prot, $i$ and the published radius $\rstar=1.45\pm0.07$~\rsun,][]{Barber2024}.  Finally, we set the respective filling factor of the small-scale field to $f_I=0.75$ -- in agreement with the results of {\tt ZeeTurbo} -- and find that the filling factor of the large-scale field that best fits our LSD Stokes $V$ profiles is equal to $f_V\simeq0.25$, slightly larger than but still consistent with that derived for other young, active low-mass stars \citep[see, e.g.,][]{Donati2023}.

The recovered large-scale magnetic topologies are dominantly poloidal, and also include a toroidal component at the stellar surface (hosting a few \% of the recovered magnetic energy) in the form of two azimuthal field rings of opposite polarities encircling the star at intermediate latitudes (in each hemisphere).  The poloidal component mainly consists of a dipole (with $\simeq$90\% of the poloidal energy) of polar strength 0.80$-$0.95~kG, slightly tilted to the rotation axis (by $\simeq$5\degr\ towards phase 0.80 in 2024 and $\simeq$15\degr\ towards phase 0.70 in 2025).  The brightness image shows only weakly contrasted features causing the observed temperature changes (see Fig.~\ref{fig:bell}) and the associated $\simeq$5\% variation in flux (at 1750~nm).  Figure~\ref{fig:mapB} shows the ZDI reconstructions from the 2024~Dec data (with the corresponding fits to the LSD Stokes $IV$ profiles and photometry displayed in Fig.~\ref{fig:fitIV}).  The other two sets of reconstructions are similar  {\emr (see Fig.~\ref{fig:mapB2})}, the main difference being the stronger dipole field and larger dipole tilt to the rotation axis in 2025~Feb, as anticipated from our \Bl\ curve whose amplitude is significantly larger at this epoch (see Fig.~\ref{fig:bell}).  We also note that the cool spot {\emr located on the magnetic equator} at phase 0.30 tends to migrate towards smaller phases, suggesting a small level of differential rotation at the surface of the star (with an equator rotating faster than the pole).  {\emr This temporal evolution over a few month is similar to what is reported for other young low-mass stars \citep[e.g.,][]{Zaire2024}}. 

\begin{figure*}[h!]
    \centering
    \includegraphics[width=0.74\textwidth]{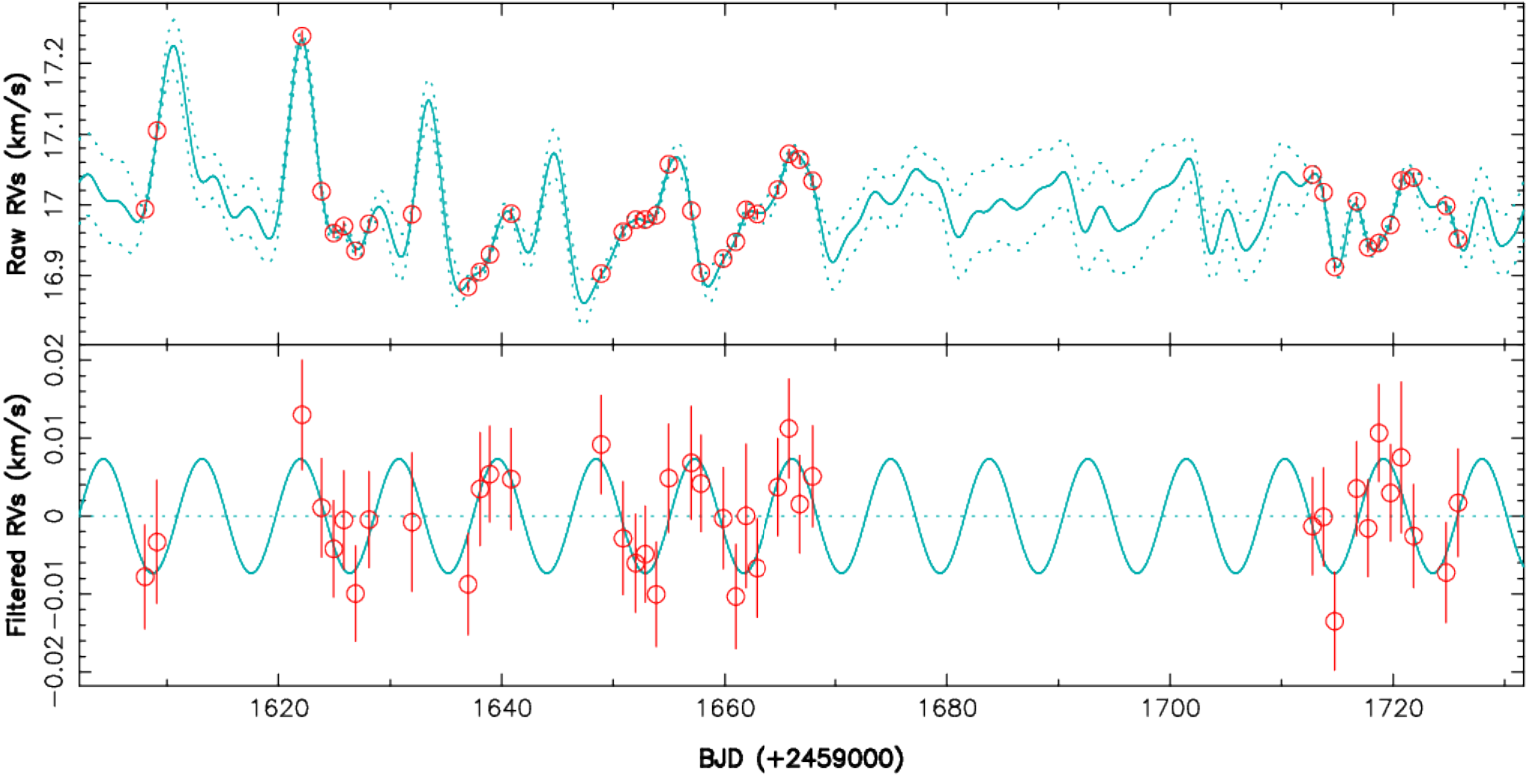}
    \caption{Raw (top) and activity-filtered (bottom) LBL RVs (red circles) derived from all spectral features of \irasstar.  The top plot shows the MCMC fit to the RV data, including GPR modeling of the activity (cyan full line, with dotted lines depicting the 68\% confidence intervals). The bottom plot shows the activity-filtered RVs with the best fit of the planet signature (cyan full line).  The rms of RV residuals is 4~\ms.}
    \label{fig:rvs}
\end{figure*}

Our modeling yields an average small-scale field of 1.4~kG in 2024 and 1.5~kG in 2025, in reasonable agreement with that estimated with {\tt ZeeTurbo} ($1.52\pm0.05$~kG).

\section{Constraining the planet mass} 
\label{sec:rvs}

To estimate the mass of \irasstar\,b, we attempted to detect the stellar barycentric motion induced by the planet.   We derived RVs from our LSD profiles of atomic lines by describing each individual profile as a first order Taylor expansion constructed from the median of all profiles \citep[as in, e.g.,][]{Donati2024}.  We proceeded in the same way to infer RVs from our LSD profiles of CO bandhead lines, which, although much sparser, are insensitive to magnetic fields and hence to activity.  All derived RVs with their uncertainties are listed in Table~\ref{tab:log}.  As is typical of low-mass stars, the CO lines are, on average, red-shifted with respect to the atomic lines (by $\simeq$0.3~\kms) due to the relative brightness of convective downwelling for the former.  

We fit these RVs with a MCMC process, using a double quasi-periodic GP model to describe the activity (one GP for each set of lines, sharing the same \Prot, hence with 9 hyper-parameters) and a planet signature assuming a circular orbit and the ephemeris from transit photometry \citep{Barber2024}.  Inferred GPR parameters are listed in Table~\ref{tab:gprv}.  As expected, atomic lines contain a 2.5$\times$ larger activity signal than the CO lines (respective average semi-amplitudes of $174\pm38$ and $69\pm23$~\ms).  The stellar rotation period we determine from the jitter is $\Prot=11.32\pm0.05$~d, falling between those inferred from \Bl\ and $dT$.  The planet signature has a derived semi-amplitude of $K=19^{+13}_{-8}$~\ms, only slightly above the 2$\sigma$ threshold.  

Being derived from all spectral features simultaneously, the LBL RVs of \irasstar\ are more precise than those inferred from LSD profiles of atomic and CO lines, with error bars in the range 6$-$13~\ms\ (see Table~\ref{tab:log}).  Fitting the LBL RVs with MCMC and GPR yields an even smaller semi-amplitude of the tentative 2$\sigma$ planet signature $K=8.6^{+9.6}_{-4.5}$~\ms, with an activity jitter of semi-amplitude $76\pm15$~\ms\ (consistent with that in the CO lines).  The resulting GPR fit is shown in Fig.~\ref{fig:rvs}, with corresponding parameters listed in Table~\ref{tab:gprv2}.  We note the significantly shorter decay time (of $18\pm4$~d), reflecting the higher RV precision and the larger activity level in the first third of our monitoring.  The corresponding periodograms, stacked periodogram and phase-folded filtered RVs are shown in Figs.~\ref{fig:per} and \ref{fig:per2}.   

The 90\% and 99\% upper limits on $K$ inferred from the MCMC modeling of the LBL RVs (and further confirmed with injection-recovery tests) are 20 and 30~\ms.    
$K=8.6^{+9.6}_{-4.5}$~\ms\ corresponds to a planet mass of $0.07^{+0.08}_{-0.04}$~\mjup, with 90\% and 99\% upper limits of 0.16 and 0.24~\mjup, respectively,  The bulk density is $0.10^{+0.11}_{-0.05}$~\gpcc, with 90\% and 99\% upper limits of 0.23 and 0.35~\gpcc, respectively (0.28 and 0.42~\gpcc, including the uncertainty on the planet radius).  

\section{Accretion/ejection from an inner disk}
\label{sec:eml}

Emission in the Paschen-$\beta$ and Brackett-$\gamma$ lines of H\,I (1282.16 and 2166.12\,nm, respectively) can be used as a measure of accretion \citep{Natta2004,Rigliaco2012}, but can also reveal outflows \citep{Whelan2004}.  There is no indication of emission in our spectra of either line but Pa-$\beta$ is slightly variable relative to neighboring lines (Fig.~\ref{fig:pabeta}).  The equivalent widths (EWs) of Pa-$\beta$ and Br-$\gamma$ are within 3.4 and 7.2~picometer (pm) respectively, of those derived from SPIRou spectra of V819~Tau and TWA~9A, young, non-accreting stars of similar spectral type.   We thus use 4 and 8~pm as upper limits for the EWs of the accretion-induced emission fluxes in these two lines, with temporal variability affecting EWs at a smaller level (see Table~\ref{tab:log} for Pa-$\beta$).  It yields maximum logarithmic accretion luminosities (in \lspy) of $-3.9\pm0.3$ using the calibrated relations of \citet{Alcala2017}, and thereby maximum logarithmic mass-accretion rates at the surface of the star $\log \dot{M}$ (in \mspy) of $-11.0\pm0.3$.  

\begin{figure}[ht!]
    \centering
    \includegraphics[width=0.8\linewidth]{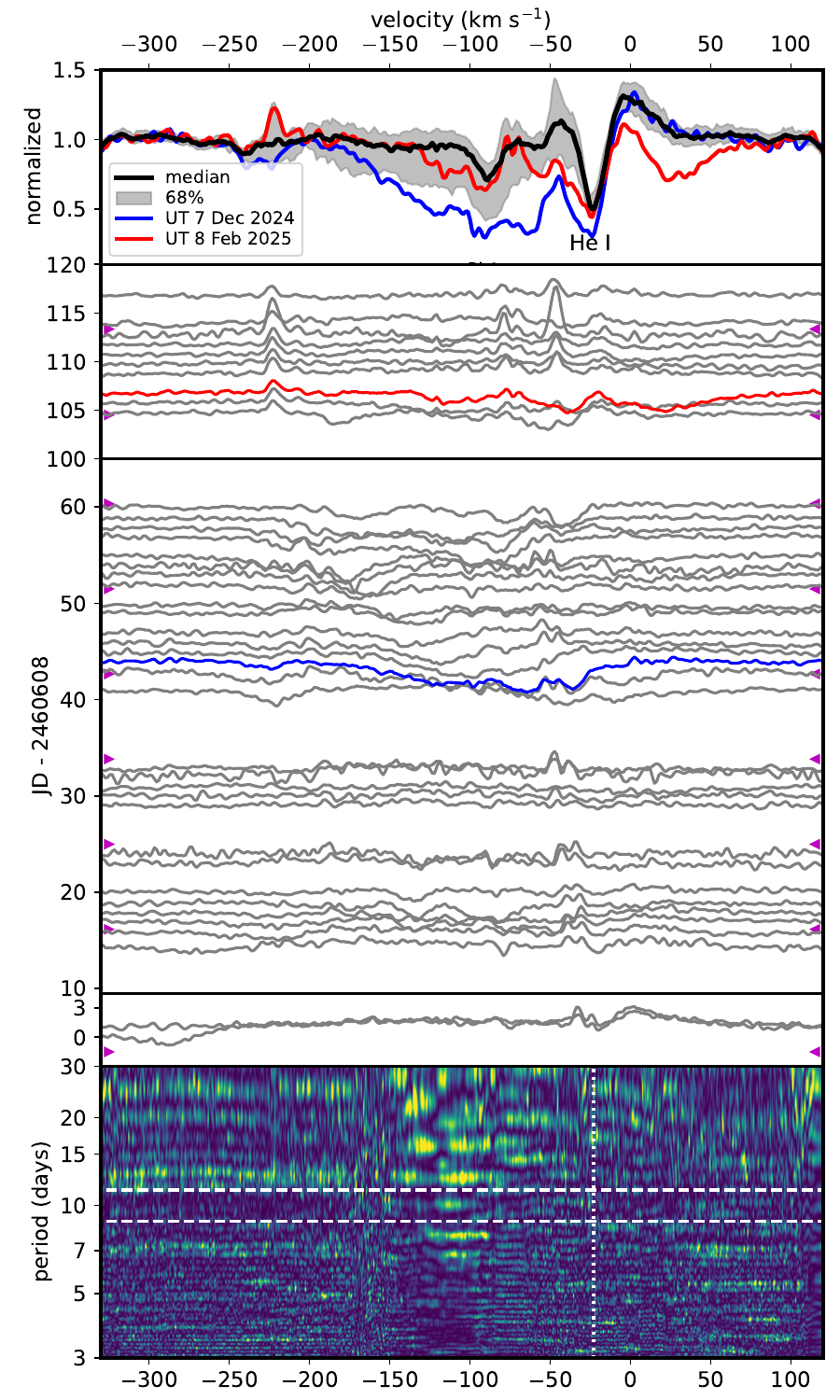}
    \caption{Top panel: median spectrum of \irasstar\ in the vicinity of the 1083-nm~\hei\ triplet, with the grey region denoting the 68\% percentile range for all spectra, and the red and blue spectra those showing the largest red-shifted (0 to 110~\kms)  and blue-shifted ($-$240 to $-$50~\kms) absorption.  Middle three panels: differential spectra relative to the median obtained at individual visits, spaced according to the relative epoch.  Note the two gaps in time.  Red and blue spectra are as in the top panel and magenta ticks mark planet transits.  Bottom panel: Lomb-Scargle periodogram, with horizontal lines marking the 8.83- and 11.3-d planet orbital and stellar rotational periods, and the vertical line marking the stable absorption feature at $-$23~\kms.}
    \label{fig:hei}
\end{figure}

Spectra of the 1083.3~nm triplet of metastable \hei\ reveal a P~Cygni-like profile with emission from the star, prominent blue-shifted absorption at $\approx-23$~\kms, and complex variability (Fig.~\ref{fig:hei}).  The red-shifted side is more stable than the blue side, but in one spectrum there is absorption (red curve in Fig. \ref{fig:hei}), suggestive of infall during magnetospheric accretion (with the Pa-$\beta$ EW at this epoch being $\simeq$0.7~\kms\ smaller than average). The velocity of the persistent absorption feature ($-$23~\kms) is much lower than that of stellar winds \citep[100s of \kms][]{Edwards2006,Kwan2007} and instead is indicative of a wind arising from an inner gaseous disk.  P Cygni-like profiles are common among the \hei\ lines of T~Tauri stars and indicate a LOS through the wind of a highly inclined disk \citep{Edwards2006,Erkal2022}.  Highly variably, blue-shifted absorption extends to $-$240~\kms, beyond which it ceases, suggesting a kinematic or chemical (\hei\ abundance) boundary  These velocities are too high for thermal/photoevaporative flow (sound speeds in H are $<12$~\kms) and instead suggest a magneto-centrifugal wind.   The spectrum with the most prominent absorption (UT 2024 Dec~07, blue curves in Fig.~\ref{fig:hei}) is the first in a series of spectra spanning two weeks showing absorption with increasing blue-shift, possibly due to structure in the wind (see Sec.~\ref{sec:discussion}). 

\section{Summary and discussion}
\label{sec:discussion}

We carried out spectropolarimetric monitoring of the planet-hosting transitional T~Tauri star \irasstar\ with SPIRou at CFHT over 118~d, securing 40 high-quality Stokes $I$ and $V$ spectra.  By applying LSD and LBL to these data, we recovered Zeeman signatures and temperature variations demonstrating that the stellar surface has variations in brightness and magnetic fields, with both \Bl\ and $dT$ modulated at \Prot\ and the modulation pattern of both quantities evolving over 118 d.  From the Zeeman broadening of magnetically sensitive lines, we infer that the small-scale surface field reaches $1.52\pm0.05$~kG. 

Our RV analysis only marginally detects the Keplerian signature of the planet, with LBL RVs yielding a semi-amplitude of $K=8.6^{+9.6}_{-4.5}$~\ms.  The corresponding planet mass is $0.07^{+0.08}_{-0.04}$~\mjup, with 90\% and 99\% upper limits of 0.16 and 0.24~\mjup, yielding a bulk density of $0.10^{+0.11}_{-0.05}$~\gpcc\ with 90\% and 99\% upper limits of 0.23 and 0.35~\gpcc\ (0.28 and 0.42~\gpcc\ including the uncertainty on the planet radius).  We rule out the scenario that \irasstar\,b is a Jovian planet and instead favor the scenario of a progenitor of a (sub) Neptune with a hydrogen-rich envelope inflated by the entropy of formation.   {\emr \irasstar\,b resembles AU~Mic~b in this respect \citep[][]{Mallorquin2024} and to a lesser extent V1298~b and c \citep[][]{Barat2024}, but not AU~Mic~c whose bulk density is much higher, demonstrating that newly formed planets need time to settle to their mature characteristics, especially those closest to their host stars.}  We can also rule out the planet as a sink for the angular momentum (AM) that would otherwise have spun up \irasstar\ since a Neptune-mass {\emr \irasstar\,b} would contain no more than 3\% of the total. 

Application of ZDI to the Stokes $I$ and $V$ LSD profiles and the photometry derived from $dT$ shows that the large-scale magnetic field of \irasstar\ is dominated by a dipole field with a strength 0.80$-$0.95~kG inclined to the rotation axis by 5\degr\ (in 2024) to 15\degr\ (in 2025).    
The star also hosts a surface toroidal field, in the form of two azimuthal field rings of opposite polarities (one in each hemisphere) encircling the star at mid-latitudes. 

\irasstar\ was identified as a possible low accretor \citep[][]{Espaillat2015}, but no excess emission appears in the Pa-$\beta$ and Br-$\gamma$ lines of H\,I, limiting accretion at the surface of the star to a maximum $\log \dot{M}$ (in \mspy) of $\simeq$$11.0\pm0.3$.  However, persistent and variable blue-shifted absorption, plus episodic red-shifted absorption in the 1083~nm \hei\ triplet unambiguously indicate the presence of an inner gaseous disk, well interior to the 20-au inner edge of the disk detected at IR/mm wavelengths,  This is not unprecedented; \citet{Thanathibodee2022} found that 30\% of a sample of disk-hosting stars without elevated H-$\alpha$ emission indicative of accretion exhibit \hei\ absorption.  

Combining the limit on $\dot{M}$  with our measurement of the large-scale magnetic dipole field in the expression of \citet{Bessolaz2008} yields a lower limit on the magnetosphere size / disk truncation radius \rmag\ of $16\pm3~\rstar$ or equivalently $\rmag\geq1.75\pm0.35~\rcor$  where \rcor\ is the corotation radius, meaning that \irasstar\ is in the ``propeller" regime of accretion \citep{Romanova2004,Ustyugova2006,Zanni2013}, with inner disk gas being expelled outwards in a magnetocentrifugal wind, presumably the one detected in the \hei\ line.   \irasstar\ shares obvious similarities with the weak propeller case of \citet{Romanova2018}, where most disc gas is ejected outwards at velocities exceeding half the escape velocity (of 300~\kms) and only a small fraction is accreted onto the star.  The ``rings" of azimuthal field could connect the star to the disk and its wind beyond \rcor\ and allow outward AM transfer, possibly maintaining the slow stellar rotation.  We speculate that accretion onto \irasstar\ was diminished by the disruption of its disk by the stellar companion, placing it prematurely into a propeller regime that has delayed its spin-up.  Its current slower rotation and concomitant lower activity have allowed, in turn the photometric detection of its transiting planet.

\begin{figure}
    \centering
    \includegraphics[width=0.9\linewidth]{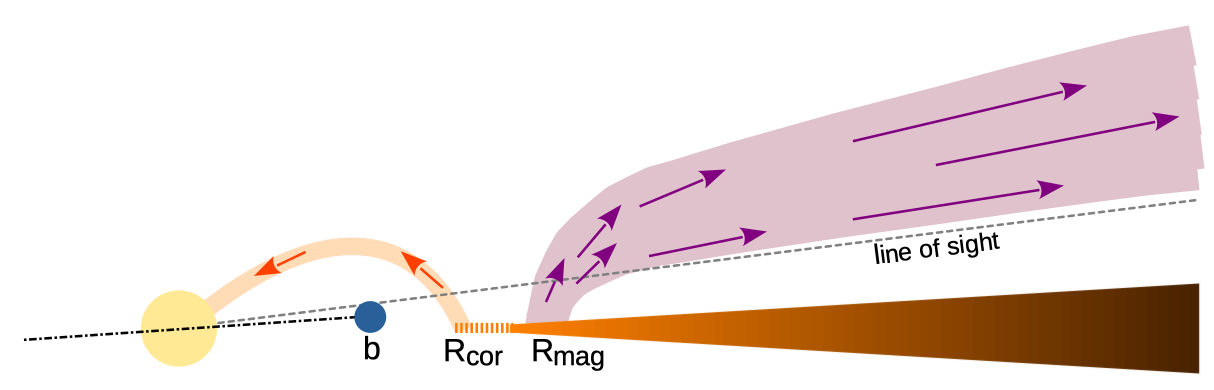}
    \caption{Illustration of the \irasstar\ system with the central star, planet (on a slightly inclined orbit), inner gaseous disk, and magnetocentrifugal wind, showing how the line of sight could produce the features in the SPIRou spectrum of the \hei\ line.}
    \label{fig:cartoon}
\end{figure}

Figure \ref{fig:cartoon} illustrates a possible geometry for the system.  An inner gas disk could survive the present level of accretion for the age of the star if its initial mass was $\gtrsim 10^{-4}$ \msun, but it could be fed from an outer disk in the manner of the PDS~70 system \citep{Gaidos2024a,Pinilla2024}.  The stable feature at $-$23~\kms\ could be produced by slower flow near the foot of the wind at low inclination to the LOS, while the variable, highly blue-shifted absorption could represent the faster region of the wind almost parallel to the LOS, with the bound to variability at $-$240~\kms\ representing the entry point of our LOS into the wind.  The tentative periodic signals in the range of periods 12$-$25~d in the absorption between $-$100 and $-$150~\kms\ (bottom panel of Fig.~\ref{fig:hei}) may correspond to Keplerian periods between \rcor\ and \rmag\ where a wind could be launched.  

More intriguing is the blue-shifted absorption event that appears after the transit of {\emr \irasstar\,b}, peaks $\sim$1~d after the transit (blue curve) and lasts about two weeks, with the peak absorption moving from about $-$40 to $-$180~\kms.  This could probe a gas over-density moving directly towards us and covering $\sim$1~au along the LOS.  Alternatively, our LOS could be passing through an extended structure, e.g., a spiral arm extending from the inner edge where velocities are slower, to further out where gas is accelerated.  Since {\emr \irasstar\,b} is probably not massive enough to gravitationally excite spiral waves \citep{Baruteau2014}, alternative mechanisms could be excitation by thermal effects \citep[shocks or shadowing]{Zhu2025} or the planet's magnetic field, or a trail of gas from the planet itself.
{With \emr \irasstar\,b located well within the stellar magnetosphere, one can also expect star-planet magnetic interactions and the generation of radio emission through the electron cyclotron maser instability \citep{Kavanagh2021}.} 

Additional RV observations could further constrain the mass of \irasstar~b.   A sub-Saturn mass and density would imply an extended low molecular weight atmosphere that is an attractive target for transit transmission spectroscopy to investigate its properties \citep[e.g., with \hst\ or \jwst,][]{Barat2024,Thao2024}.  Long-term spectropolarimetric monitoring with SPIRou could reveal an evolution in the strength and geometry of the magnetic field as has been seen on other T~Tauri stars \citep[e.g.,][]{Donati2024} and, via the \hei line, the response of the inner disk and its wind to these changes.  Observations at higher cadence could search for a repetition of the accelerating blue-shifted absorption transient and evaluate whether there is a potential connection with the planet.  More sensitive searches for inner disk gas (e.g., in CO lines with ALMA) and dust (with \jwst) are also warranted.  Finally, radio observations of \irasstar\ with, e.g., LOFAR, to detect and characterize star-planet magnetic interactions \citep{Kavanagh2021} could ultimately allow us to constrain the planet's magnetic field.  

\begin{acknowledgements}
EG and AM were supported by NASA Awards 80NSSC20K0957 and 80NSSC25K7148 (Exoplanets Research Program).  Our study is based on data obtained at the CFHT, operated by the CNRC (Canada), INSU/CNRS (France) and the University of Hawai'i.  
\end{acknowledgements}

\bibliographystyle{aa} 
\bibliography{aa54628-25} 

\begin{thebibliography}{51}
\expandafter\ifx\csname natexlab\endcsname\relax\def\natexlab#1{#1}\fi

\bibitem[{{Alcal{\'a}} {et~al.}(2017){Alcal{\'a}}, {Manara}, {Natta}, {Frasca},
  {Testi}, {Nisini}, {Stelzer}, {Williams}, {Antoniucci}, {Biazzo}, {Covino},
  {Esposito}, {Getman}, \& {Rigliaco}}]{Alcala2017}
{Alcal{\'a}}, J.~M., {Manara}, C.~F., {Natta}, A., {et~al.} 2017, \aap, 600,
  A20

\bibitem[{{Artigau} {et~al.}(2024){Artigau}, {Cadieux}, {Cook}, {Doyon},
  {Dauplaise}, {Arnold}, {Cadieux}, {Donati}, {Cristofari}, {Delfosse},
  {Fouqu{\'e}}, {Moutou}, {Larue}, \& {Allart}}]{Artigau2024}
{Artigau}, {\'E}., {Cadieux}, C., {Cook}, N.~J., {et~al.} 2024, \aj, 168, 252

\bibitem[{{Artigau} {et~al.}(2022){Artigau}, {Cadieux}, {Cook}, {Doyon},
  {Vandal}, {Donati}, {Moutou}, {Delfosse}, {Fouqu{\'e}}, {Martioli}, {Bouchy},
  {Parsons}, {Carmona}, {Dumusque}, {Astudillo-Defru}, {Bonfils}, \&
  {Mignon}}]{Artigau2022}
{Artigau}, {\'E}., {Cadieux}, C., {Cook}, N.~J., {et~al.} 2022, \aj, 164, 84

\bibitem[{{Barat} {et~al.}(2024){Barat}, {D{\'e}sert}, {Goyal}, {Vazan},
  {Kawashima}, {Fortney}, {Bean}, {Line}, {Panwar}, {Jacobs}, {Shivkumar},
  {Sikora}, {Baeyens}, {Oklop{\v{c}}i{\'c}}, {David}, \&
  {Livingston}}]{Barat2024}
{Barat}, S., {D{\'e}sert}, J.-M., {Goyal}, J.~M., {et~al.} 2024, \aap, 692,
  A198

\bibitem[{{Barber} {et~al.}(2024{\natexlab{a}}){Barber}, {Mann}, {Vanderburg},
  {Krolikowski}, {Kraus}, {Ansdell}, {Pearce}, {Mace}, {Andrews}, {Boyle},
  {Collins}, {De Furio}, {Dragomir}, {Espaillat}, {Feinstein}, {Fields},
  {Jaffe}, {Lopez Murillo}, {Murgas}, {Newton}, {Palle}, {Sawczynec},
  {Schwarz}, {Thao}, {Tofflemire}, {Watkins}, {Jenkins}, {Latham}, {Ricker},
  {Seager}, {Vanderspek}, {Winn}, {Charbonneau}, {Essack}, {Rodriguez},
  {Shporer}, {Twicken}, \& {Villase{\~n}or}}]{Barber2024}
{Barber}, M.~G., {Mann}, A.~W., {Vanderburg}, A., {et~al.} 2024{\natexlab{a}},
  \nat, 635, 574

\bibitem[{{Barber} {et~al.}(2024{\natexlab{b}}){Barber}, {Thao}, {Mann},
  {Vanderburg}, {Mori}, {Livingston}, {Fukui}, {Narita}, {Kraus}, {Tofflemire},
  {Newton}, {Winn}, {Jenkins}, {Seager}, {Collins}, \& {Twicken}}]{Barber2024b}
{Barber}, M.~G., {Thao}, P.~C., {Mann}, A.~W., {et~al.} 2024{\natexlab{b}},
  \apjl, 973, L30

\bibitem[{{Baruteau} {et~al.}(2014){Baruteau}, {Crida}, {Paardekooper},
  {Masset}, {Guilet}, {Bitsch}, {Nelson}, {Kley}, \&
  {Papaloizou}}]{Baruteau2014}
{Baruteau}, C., {Crida}, A., {Paardekooper}, S.~J., {et~al.} 2014, in
  Protostars and Planets VI, ed. H.~{Beuther}, R.~S. {Klessen}, C.~P.
  {Dullemond}, \& T.~{Henning}, 667--689

\bibitem[{{Bessolaz} {et~al.}(2008){Bessolaz}, {Zanni}, {Ferreira}, {Keppens},
  \& {Bouvier}}]{Bessolaz2008}
{Bessolaz}, N., {Zanni}, C., {Ferreira}, J., {Keppens}, R., \& {Bouvier}, J.
  2008, \aap, 478, 155

\bibitem[{{Chib} \& {Jeliazkov}(2001)}]{Chib2001}
{Chib}, S. \& {Jeliazkov}, I. 2001, Journal of the American Statistical
  Association, 96, 270

\bibitem[{{Cook} {et~al.}(2022){Cook}, {Artigau}, {Doyon}, {Hobson},
  {Martioli}, {Bouchy}, {Moutou}, {Carmona}, {Usher}, {Fouqu{\'e}}, {Arnold},
  {Delfosse}, {Boisse}, {Cadieux}, {Vandal}, {Donati}, \&
  {Desli{\`e}res}}]{Cook2022}
{Cook}, N.~J., {Artigau}, {\'E}., {Doyon}, R., {et~al.} 2022, \pasp, 134,
  114509

\bibitem[{{Cristofari} {et~al.}(2023){Cristofari}, {Donati}, {Folsom},
  {Masseron}, {Fouqu{\'e}}, {Moutou}, {Artigau}, {Carmona}, {Petit},
  {Delfosse}, {Martioli}, \& {the SLS consortium}}]{Cristofari2023}
{Cristofari}, P.~I., {Donati}, J.~F., {Folsom}, C.~P., {et~al.} 2023, \mnras,
  522, 1342

\bibitem[{{Dai} {et~al.}(2024){Dai}, {Goldberg}, {Batygin}, {van Saders},
  {Chiang}, {Choksi}, {Li}, {Petigura}, {Gilbert}, {Millholland}, {Dai},
  {Bouma}, {Weiss}, \& {Winn}}]{Dai2024}
{Dai}, F., {Goldberg}, M., {Batygin}, K., {et~al.} 2024, \aj, 168, 239

\bibitem[{{Donati} {et~al.}(2024){Donati}, {Cristofari}, {Alencar},
  {K{\'o}sp{\'a}l}, {Bouvier}, {Moutou}, {Carmona}, {Gregorio-Hetem},
  {M{\'e}nard}, {Artigau}, {Doyon}, {Takami}, {Shang}, {Dias do Nascimento},
  {M{\'e}nard}, {Gaidos}, \& {SPIRou Science Team}}]{Donati2024}
{Donati}, J.~F., {Cristofari}, P.~I., {Alencar}, S.~H.~P., {et~al.} 2024,
  \mnras, 535, 3363

\bibitem[{{Donati} {et~al.}(2023){Donati}, {Cristofari}, {Finociety}, {Klein},
  {Moutou}, {Gaidos}, {Cadieux}, {Artigau}, {Correia}, {Bou{\'e}}, {Cook},
  {Carmona}, {Lehmann}, {Bouvier}, {Martioli}, {Morin}, {Fouqu{\'e}},
  {Delfosse}, {Doyon}, {H{\'e}brard}, {Alencar}, {Laskar}, {Arnold}, {Petit},
  {K{\'o}sp{\'a}l}, {Vidotto}, {Folsom}, \& {collaboration}}]{Donati2023}
{Donati}, J.~F., {Cristofari}, P.~I., {Finociety}, B., {et~al.} 2023, \mnras,
  525, 455

\bibitem[{{Donati} {et~al.}(2006){Donati}, {Howarth}, {Jardine}, {Petit},
  {Catala}, {Landstreet}, {Bouret}, {Alecian}, {Barnes}, {Forveille},
  {Paletou}, \& {Manset}}]{Donati2006}
{Donati}, J.-F., {Howarth}, I.~D., {Jardine}, M.~M., {et~al.} 2006, \mnras,
  370, 629

\bibitem[{{Donati} {et~al.}(2020){Donati}, {Kouach}, {Moutou}, {Doyon},
  {Delfosse}, {Artigau}, {Baratchart}, {Lacombe}, {Barrick}, {H{\'e}brard},
  {Bouchy}, {Saddlemyer}, {Par{\`e}s}, {Rabou}, {Micheau}, {Dolon}, {Reshetov},
  {Challita}, {Carmona}, {Striebig}, {Thibault}, {Martioli}, {Cook},
  {Fouqu{\'e}}, {Vermeulen}, {Wang}, {Arnold}, {Pepe}, {Boisse}, {Figueira},
  {Bouvier}, {Ray}, {Feugeade}, {Morin}, {Alencar}, {Hobson}, {Castilho},
  {Udry}, {Santos}, {Hernandez}, {Benedict}, {Vall{\'e}e}, {Gallou}, {Dupieux},
  {Larrieu}, {Perruchot}, {Sottile}, {Moreau}, {Usher}, {Baril}, {Wildi},
  {Chazelas}, {Malo}, {Bonfils}, {Loop}, {Kerley}, {Wevers}, {Dunn}, {Pazder},
  {Macdonald}, {Dubois}, {Carri{\'e}}, {Valentin}, {Henault}, {Yan}, \&
  {Steinmetz}}]{Donati2020}
{Donati}, J.~F., {Kouach}, D., {Moutou}, C., {et~al.} 2020, \mnras, 498, 5684

\bibitem[{{Donati} {et~al.}(1997){Donati}, {Semel}, {Carter}, {Rees}, \&
  {Collier Cameron}}]{Donati1997}
{Donati}, J.-F., {Semel}, M., {Carter}, B.~D., {Rees}, D.~E., \& {Collier
  Cameron}, A. 1997, \mnras, 291, 658

\bibitem[{{Edwards} {et~al.}(2006){Edwards}, {Fischer}, {Hillenbrand}, \&
  {Kwan}}]{Edwards2006}
{Edwards}, S., {Fischer}, W., {Hillenbrand}, L., \& {Kwan}, J. 2006, \apj, 646,
  319

\bibitem[{{Erkal} {et~al.}(2022){Erkal}, {Manara}, {Schneider}, {Vincenzi},
  {Nisini}, {Coffey}, {Alcal{\'a}}, {Fedele}, \& {Antoniucci}}]{Erkal2022}
{Erkal}, J., {Manara}, C.~F., {Schneider}, P.~C., {et~al.} 2022, \aap, 666,
  A188

\bibitem[{{Espaillat} {et~al.}(2015){Espaillat}, {Andrews}, {Powell},
  {Feldman}, {Qi}, {Wilner}, \& {D'Alessio}}]{Espaillat2015}
{Espaillat}, C., {Andrews}, S., {Powell}, D., {et~al.} 2015, \apj, 807, 156

\bibitem[{{Fernandes} {et~al.}(2025){Fernandes}, {Bergsten}, {Mulders},
  {Pascucci}, {Hardegree-Ullman}, {Giacalone}, {Christiansen}, {Rogers},
  {Gupta}, {Dawson}, {Koskinen}, {Boley}, {Curtis}, {Cunha}, {Mamajek},
  {Sagynbayeva}, {Bhure}, {Ciardi}, {Karpoor}, {Pearson}, {Zink}, \&
  {Feiden}}]{Fernandes2025}
{Fernandes}, R.~B., {Bergsten}, G.~J., {Mulders}, G.~D., {et~al.} 2025, \aj,
  169, 208

\bibitem[{{Finociety} \& {Donati}(2022)}]{Finociety2022}
{Finociety}, B. \& {Donati}, J.~F. 2022, \mnras, 516, 5887

\bibitem[{{Furlan} {et~al.}(2009){Furlan}, {Watson}, {McClure}, {Manoj},
  {Espaillat}, {D'Alessio}, {Calvet}, {Kim}, {Sargent}, {Forrest}, \&
  {Hartmann}}]{Furlan2009}
{Furlan}, E., {Watson}, D.~M., {McClure}, M.~K., {et~al.} 2009, \apj, 703, 1964

\bibitem[{{Gaidos} {et~al.}(2024){Gaidos}, {Thanathibodee}, {Hoffman}, {Ong},
  {Hinkle}, {Shappee}, \& {Banzatti}}]{Gaidos2024a}
{Gaidos}, E., {Thanathibodee}, T., {Hoffman}, A., {et~al.} 2024, \apj, 966, 167

\bibitem[{{Gehrig} {et~al.}(2023){Gehrig}, {Gaidos}, \&
  {G{\"u}del}}]{Gehrig2023}
{Gehrig}, L., {Gaidos}, E., \& {G{\"u}del}, M. 2023, \aap, 675, A179

\bibitem[{{Green}(2018)}]{Green2018}
{Green}, G. 2018, The Journal of Open Source Software, 3, 695

\bibitem[{{Haywood} {et~al.}(2014){Haywood}, {Collier Cameron}, {Queloz},
  {Barros}, {Deleuil}, {Fares}, {Gillon}, {Lanza}, {Lovis}, {Moutou}, {Pepe},
  {Pollacco}, {Santerne}, {S{\'e}gransan}, \& {Unruh}}]{Haywood2014}
{Haywood}, R.~D., {Collier Cameron}, A., {Queloz}, D., {et~al.} 2014, \mnras,
  443, 2517

\bibitem[{{Kavanagh} {et~al.}(2021){Kavanagh}, {Vidotto}, {Klein}, {Jardine},
  {Donati}, \& {{\'O} Fionnag{\'a}in}}]{Kavanagh2021}
{Kavanagh}, R.~D., {Vidotto}, A.~A., {Klein}, B., {et~al.} 2021, \mnras, 504,
  1511

\bibitem[{{Kwan} {et~al.}(2007){Kwan}, {Edwards}, \& {Fischer}}]{Kwan2007}
{Kwan}, J., {Edwards}, S., \& {Fischer}, W. 2007, \apj, 657, 897

\bibitem[{{Landi degl'Innocenti} \& {Landolfi}(2004)}]{Landi2004}
{Landi degl'Innocenti}, E. \& {Landolfi}, M. 2004, {Polarisation in spectral
  lines} (Dordrecht/Boston/London: Kluwer Academic Publishers)

\bibitem[{{L{\'o}pez-Valdivia} {et~al.}(2021){L{\'o}pez-Valdivia}, {Sokal},
  {Mace}, {Kidder}, {Hussaini}, {Nofi}, {Prato}, {Johns-Krull}, {Oh}, {Lee},
  {Park}, {Oh}, {Kraus}, {Kaplan}, {Llama}, {Mann}, {Kim}, {Gully-Santiago},
  {Lee}, {Pak}, {Hwang}, \& {Jaffe}}]{Lopez-Valdivia2021}
{L{\'o}pez-Valdivia}, R., {Sokal}, K.~R., {Mace}, G.~N., {et~al.} 2021, \apj,
  921, 53

\bibitem[{{Luhman}(2025)}]{Luhman2025}
{Luhman}, K.~L. 2025, \aj, 169, 179

\bibitem[{{Luhman} {et~al.}(2009){Luhman}, {Mamajek}, {Allen}, {Muench}, \&
  {Finkbeiner}}]{Luhman2009}
{Luhman}, K.~L., {Mamajek}, E.~E., {Allen}, P.~R., {Muench}, A.~A., \&
  {Finkbeiner}, D.~P. 2009, \apj, 691, 1265

\bibitem[{{Mallorqu{\'\i}n} {et~al.}(2024){Mallorqu{\'\i}n}, {B{\'e}jar},
  {Lodieu}, {Zapatero Osorio}, {Yu}, {Su{\'a}rez Mascare{\~n}o}, {Damasso},
  {Sanz-Forcada}, {Ribas}, {Reiners}, {Quirrenbach}, {Amado}, {Caballero},
  {Aigrain}, {Barrag{\'a}n}, {Dreizler}, {Fern{\'a}ndez-Mart{\'\i}n}, {Goffo},
  {Henning}, {Kaminski}, {Klein}, {Luque}, {Montes}, {Morales}, {Nagel},
  {Pall{\'e}}, {Reffert}, {Schlecker}, \& {Schweitzer}}]{Mallorquin2024}
{Mallorqu{\'\i}n}, M., {B{\'e}jar}, V.~J.~S., {Lodieu}, N., {et~al.} 2024,
  \aap, 689, A132

\bibitem[{{Natta} {et~al.}(2004){Natta}, {Testi}, {Muzerolle}, {Randich},
  {Comer{\'o}n}, \& {Persi}}]{Natta2004}
{Natta}, A., {Testi}, L., {Muzerolle}, J., {et~al.} 2004, \aap, 424, 603

\bibitem[{{Pinilla} {et~al.}(2024){Pinilla}, {Benisty}, {Waters}, {Bae}, \&
  {Facchini}}]{Pinilla2024}
{Pinilla}, P., {Benisty}, M., {Waters}, R., {Bae}, J., \& {Facchini}, S. 2024,
  \aap, 686, A135

\bibitem[{{Rajpaul} {et~al.}(2015){Rajpaul}, {Aigrain}, {Osborne}, {Reece}, \&
  {Roberts}}]{Rajpaul2015}
{Rajpaul}, V., {Aigrain}, S., {Osborne}, M.~A., {Reece}, S., \& {Roberts}, S.
  2015, \mnras, 452, 2269

\bibitem[{{Rebull} {et~al.}(2020){Rebull}, {Stauffer}, {Cody}, {Hillenbrand},
  {Bouvier}, {Roggero}, \& {David}}]{Rebull2020}
{Rebull}, L.~M., {Stauffer}, J.~R., {Cody}, A.~M., {et~al.} 2020, \aj, 159, 273

\bibitem[{{Rigliaco} {et~al.}(2012){Rigliaco}, {Natta}, {Testi}, {Randich},
  {Alcal{\`a}}, {Covino}, \& {Stelzer}}]{Rigliaco2012}
{Rigliaco}, E., {Natta}, A., {Testi}, L., {et~al.} 2012, \aap, 548, A56

\bibitem[{{Romanova} {et~al.}(2018){Romanova}, {Blinova}, {Ustyugova},
  {Koldoba}, \& {Lovelace}}]{Romanova2018}
{Romanova}, M.~M., {Blinova}, A.~A., {Ustyugova}, G.~V., {Koldoba}, A.~V., \&
  {Lovelace}, R.~V.~E. 2018, \na, 62, 94

\bibitem[{{Romanova} {et~al.}(2004){Romanova}, {Ustyugova}, {Koldoba}, \&
  {Lovelace}}]{Romanova2004}
{Romanova}, M.~M., {Ustyugova}, G.~V., {Koldoba}, A.~V., \& {Lovelace},
  R.~V.~E. 2004, \apjl, 616, L151

\bibitem[{{Ryabchikova} {et~al.}(2015){Ryabchikova}, {Piskunov}, {Kurucz},
  {Stempels}, {Heiter}, {Pakhomov}, \& {Barklem}}]{Ryabchikova2015}
{Ryabchikova}, T., {Piskunov}, N., {Kurucz}, R.~L., {et~al.} 2015, \physscr,
  90, 054005

\bibitem[{{Serna} {et~al.}(2021){Serna}, {Hernandez}, {Kounkel},
  {Manzo-Mart{\'\i}nez}, {Roman-Lopes}, {Rom{\'a}n-Z{\'u}{\~n}iga}, {Batista},
  {Pinz{\'o}n}, {Calvet}, {Brice{\~n}o}, {Tapia}, {Su{\'a}rez}, {Ram{\'\i}rez},
  {G. Stassun}, {Covey}, {Vargas-Gonz{\'a}lez}, \&
  {Fern{\'a}ndez-Trincado}}]{Serna2021}
{Serna}, J., {Hernandez}, J., {Kounkel}, M., {et~al.} 2021, \apj, 923, 177

\bibitem[{{Skilling} \& {Bryan}(1984)}]{Skilling1984}
{Skilling}, J. \& {Bryan}, R.~K. 1984, \mnras, 211, 111

\bibitem[{{Thanathibodee} {et~al.}(2022){Thanathibodee}, {Calvet},
  {Hern{\'a}ndez}, {Mauc{\'o}}, \& {Brice{\~n}o}}]{Thanathibodee2022}
{Thanathibodee}, T., {Calvet}, N., {Hern{\'a}ndez}, J., {Mauc{\'o}}, K., \&
  {Brice{\~n}o}, C. 2022, \aj, 163, 74

\bibitem[{{Thao} {et~al.}(2024){Thao}, {Mann}, {Feinstein}, {Gao}, {Thorngren},
  {Rotman}, {Welbanks}, {Brown}, {Duvvuri}, {France}, {Longo}, {Sandoval},
  {Schneider}, {Wilson}, {Youngblood}, {Vanderburg}, {Barber}, {Wood},
  {Batalha}, {Kraus}, {Murray}, {Newton}, {Rizzuto}, {Tofflemire}, {Tsai},
  {Bean}, {Berta-Thompson}, {Evans-Soma}, {Froning}, {Kempton}, {Miguel}, \&
  {Pineda}}]{Thao2024}
{Thao}, P.~C., {Mann}, A.~W., {Feinstein}, A.~D., {et~al.} 2024, \aj, 168, 297

\bibitem[{{Ustyugova} {et~al.}(2006){Ustyugova}, {Koldoba}, {Romanova}, \&
  {Lovelace}}]{Ustyugova2006}
{Ustyugova}, G.~V., {Koldoba}, A.~V., {Romanova}, M.~M., \& {Lovelace},
  R.~V.~E. 2006, \apj, 646, 304

\bibitem[{{Whelan} {et~al.}(2004){Whelan}, {Ray}, \& {Davis}}]{Whelan2004}
{Whelan}, E.~T., {Ray}, T.~P., \& {Davis}, C.~J. 2004, \aap, 417, 247

\bibitem[{{Zaire} {et~al.}(2024){Zaire}, {Donati}, {Alencar}, {Bouvier},
  {Moutou}, {Bellotti}, {Carmona}, {Petit}, {K{\'o}sp{\'a}l}, {Shang},
  {Grankin}, {Manara}, {Alecian}, {Gregory}, {Fouqu{\'e}}, \& {the SLS
  consortium}}]{Zaire2024}
{Zaire}, B., {Donati}, J.~F., {Alencar}, S.~P., {et~al.} 2024, \mnras, 533,
  2893

\bibitem[{{Zanni} \& {Ferreira}(2013)}]{Zanni2013}
{Zanni}, C. \& {Ferreira}, J. 2013, \aap, 550, A99

\bibitem[{{Zhu} {et~al.}(2025){Zhu}, {Zhang}, \& {Johnson}}]{Zhu2025}
{Zhu}, Z., {Zhang}, S., \& {Johnson}, T.~M. 2025, \apj, 980, 259

\end{thebibliography}

%

\begin{appendix} 

\section{TESS light curve during sector 86}
\label{sec:app}

\begin{figure}
    \centering
    \includegraphics[width=\linewidth]{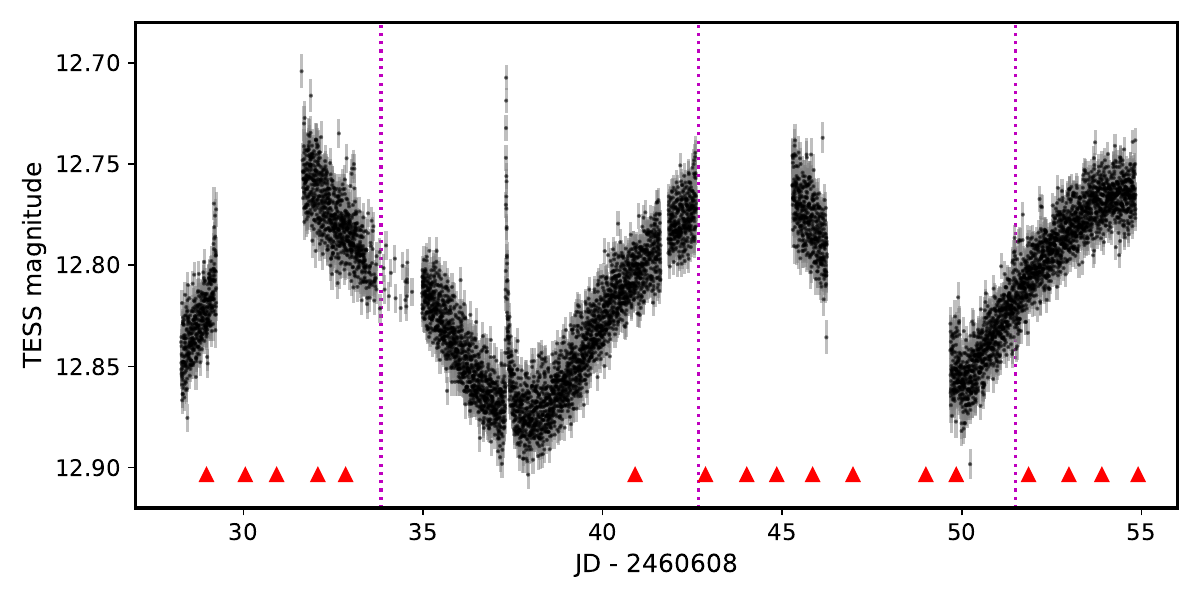}
    \caption{200-sec cadence \tess\ light curve of \irasstar\ during Sector 86, as obtained by {\tt TESSExtractor} \citep{Serna2021}, showing rotational variability and a single flare.  Red triangles indicate epochs of SPIRou observations and vertical dotted lines indicate transits of the planet {\emr \irasstar\,b}.}
    \label{fig:tess}
\end{figure}

\section{Observation log}
\label{sec:appA}

Table~\ref{tab:log} gives the full log and associated \Bl, $dT$, RV and Pa-$\beta$ EW measurements at each observing epoch from our SPIRou spectra of \irasstar.

\begin{table*} 
\small
\caption[]{Log of our SPIRou observations of \irasstar\ in semesters 2024B and 2025A.} 
\centering                      
\resizebox{0.95\textwidth}{!}{
\begin{tabular}{ccccccccccccc}
\hline
BJD        & UT date & c / $\phi$ & o / $\psi$ & t$_{\rm exp}$ & SNR  & $\sigma_V$      & \Bl\  &  $dT$ & RV (ato)   & RV (CO) & RV (LBL) & Pa-$\beta$ EW \\
(2460000+) &         & (rotation) & (orbit)   &   (s)         & ($H$)& ($10^{-4} I_c$) & (G)   &  (K)  & (\kms) & (\kms) & (\kms) & (\kms)  \\
\hline
608.0143940 & 2024 Oct 24 & 0 / 0.001 & 202 / 0.172 & 2128.5 & 147 & 4.3 & 24.4$\pm$8.8 & 10.0$\pm$1.4 & 16.69$\pm$0.05 & 17.13$\pm$0.05 & 16.994$\pm$0.007 & -4.3$\pm$0.2 \\
609.0920572 & 2024 Oct 25 & 0 / 0.096 & 202 / 0.294 & 2407.1 & 119 & 6.1 & 61.4$\pm$12.6 & 14.8$\pm$1.7 & 16.70$\pm$0.06 & 17.06$\pm$0.08 & 17.105$\pm$0.008 & -5.6$\pm$0.2 \\
622.0988784 & 2024 Nov 07 & 1 / 0.242 & 203 / 0.767 & 2407.1 & 124 & 5.3 & 51.4$\pm$11.4 & -41.3$\pm$1.6 & 16.86$\pm$0.05 & 17.09$\pm$0.07 & 17.239$\pm$0.007 & -5.5$\pm$0.2 \\
623.8468610 & 2024 Nov 09 & 1 / 0.396 & 203 / 0.964 & 2351.4 & 144 & 4.5 & 20.8$\pm$10.1 & -49.2$\pm$1.3 & 16.45$\pm$0.05 & 16.93$\pm$0.06 & 17.019$\pm$0.006 & -6.5$\pm$0.2 \\
624.9390119 & 2024 Nov 10 & 1 / 0.492 & 204 / 0.088 & 2039.3 & 147 & 4.3 & 44.0$\pm$8.9 & -22.5$\pm$1.3 & 16.41$\pm$0.05 & 16.95$\pm$0.05 & 16.960$\pm$0.006 & -6.0$\pm$0.2 \\
625.8439799 & 2024 Nov 11 & 1 / 0.572 & 204 / 0.191 & 2044.9 & 147 & 4.2 & 16.2$\pm$9.4 & -4.5$\pm$1.3 & 16.30$\pm$0.06 & 16.86$\pm$0.06 & 16.970$\pm$0.006 & -6.7$\pm$0.2 \\
626.8787366 & 2024 Nov 12 & 1 / 0.663 & 204 / 0.308 & 2078.3 & 146 & 4.2 & 0.3$\pm$9.1 & 18.2$\pm$1.3 & 16.44$\pm$0.05 & 16.91$\pm$0.06 & 16.935$\pm$0.006 & -6.8$\pm$0.2 \\
628.1004148 & 2024 Nov 13 & 1 / 0.771 & 204 / 0.446 & 2395.9 & 145 & 4.6 & 10.8$\pm$9.3 & 23.3$\pm$1.3 & 16.71$\pm$0.05 & 17.03$\pm$0.06 & 16.973$\pm$0.006 & -5.5$\pm$0.2 \\
631.9550217 & 2024 Nov 17 & 2 / 0.111 & 204 / 0.882 & 2407.1 & 92 & 7.9 & 66.2$\pm$17.7 & -1.5$\pm$2.0 & 16.79$\pm$0.07 & 17.06$\pm$0.08 & 16.986$\pm$0.009 & -6.2$\pm$0.2 \\
636.9760053 & 2024 Nov 22 & 2 / 0.553 & 205 / 0.451 & 2407.1 & 132 & 5.1 & 23.8$\pm$10.5 & -5.4$\pm$1.4 & 16.55$\pm$0.05 & 16.88$\pm$0.06 & 16.884$\pm$0.006 & -6.6$\pm$0.2 \\
638.0567321 & 2024 Nov 23 & 2 / 0.648 & 205 / 0.573 & 2407.1 & 121 & 6.3 & 19.2$\pm$13.2 & 14.3$\pm$1.6 & 16.59$\pm$0.05 & 16.99$\pm$0.07 & 16.905$\pm$0.007 & -6.8$\pm$0.2 \\
638.9277689 & 2024 Nov 24 & 2 / 0.725 & 205 / 0.671 & 2301.2 & 144 & 4.5 & 37.9$\pm$9.2 & 18.9$\pm$1.3 & 16.73$\pm$0.05 & 16.94$\pm$0.05 & 16.929$\pm$0.006 & -6.3$\pm$0.2 \\
640.8493921 & 2024 Nov 26 & 2 / 0.894 & 205 / 0.889 & 2368.1 & 139 & 5.0 & 16.9$\pm$10.6 & 14.5$\pm$1.4 & 16.81$\pm$0.05 & 17.05$\pm$0.05 & 16.987$\pm$0.006 & -5.9$\pm$0.2 \\
648.9055747 & 2024 Dec 04 & 3 / 0.604 & 206 / 0.801 & 2100.6 & 137 & 4.6 & 29.2$\pm$9.4 & 5.9$\pm$1.3 & 16.61$\pm$0.05 & 16.94$\pm$0.06 & 16.902$\pm$0.006 & -6.0$\pm$0.2 \\
650.8658794 & 2024 Dec 06 & 3 / 0.777 & 207 / 0.023 & 2407.1 & 116 & 5.8 & 55.5$\pm$12.2 & 23.5$\pm$1.6 & 16.63$\pm$0.06 & 17.01$\pm$0.07 & 16.961$\pm$0.007 & -6.0$\pm$0.2 \\
652.0129438 & 2024 Dec 07 & 3 / 0.878 & 207 / 0.152 & 2407.1 & 135 & 4.8 & 27.1$\pm$10.2 & 15.8$\pm$1.3 & 16.78$\pm$0.05 & 17.06$\pm$0.05 & 16.979$\pm$0.006 & -6.3$\pm$0.2 \\
652.8490028 & 2024 Dec 08 & 3 / 0.951 & 207 / 0.247 & 2234.3 & 146 & 4.4 & -4.1$\pm$8.9 & 11.0$\pm$1.3 & 16.78$\pm$0.05 & 17.06$\pm$0.05 & 16.979$\pm$0.006 & -5.5$\pm$0.2 \\
653.8467623 & 2024 Dec 09 & 4 / 0.039 & 207 / 0.360 & 2245.5 & 130 & 4.9 & -0.8$\pm$10.6 & 3.7$\pm$1.4 & 16.83$\pm$0.06 & 17.05$\pm$0.06 & 16.985$\pm$0.007 & -6.2$\pm$0.2 \\
654.9702164 & 2024 Dec 10 & 4 / 0.138 & 207 / 0.487 & 2407.1 & 124 & 5.4 & 58.7$\pm$11.3 & -20.1$\pm$1.5 & 16.94$\pm$0.05 & 17.12$\pm$0.07 & 17.057$\pm$0.007 & -5.7$\pm$0.2 \\
656.9989921 & 2024 Dec 12 & 4 / 0.317 & 207 / 0.717 & 2407.1 & 125 & 5.2 & 16.7$\pm$11.5 & -62.6$\pm$1.6 & 16.66$\pm$0.06 & 17.05$\pm$0.06 & 16.991$\pm$0.007 & -6.8$\pm$0.2 \\
657.8461617 & 2024 Dec 13 & 4 / 0.392 & 207 / 0.813 & 2245.5 & 138 & 4.6 & 38.1$\pm$9.9 & -47.5$\pm$1.3 & 16.44$\pm$0.05 & 16.89$\pm$0.04 & 16.904$\pm$0.006 & -6.9$\pm$0.2 \\
659.8565597 & 2024 Dec 15 & 4 / 0.569 & 208 / 0.040 & 2295.6 & 130 & 5.2 & 50.0$\pm$11.0 & -4.5$\pm$1.4 & 16.57$\pm$0.05 & 16.93$\pm$0.07 & 16.924$\pm$0.006 & -6.2$\pm$0.2 \\
660.9820562 & 2024 Dec 16 & 4 / 0.668 & 208 / 0.168 & 2373.6 & 128 & 5.2 & 29.6$\pm$10.7 & 10.8$\pm$1.4 & 16.61$\pm$0.05 & 16.94$\pm$0.05 & 16.947$\pm$0.007 & -6.9$\pm$0.2 \\
661.8967913 & 2024 Dec 17 & 4 / 0.749 & 208 / 0.271 & 2407.1 & 89 & 8.2 & 59.4$\pm$18.1 & 24.1$\pm$2.1 & 16.71$\pm$0.11 & 16.94$\pm$0.08 & 16.993$\pm$0.009 & -7.0$\pm$0.2 \\
662.9070342 & 2024 Dec 18 & 4 / 0.838 & 208 / 0.386 & 2401.5 & 141 & 4.6 & 39.8$\pm$9.3 & 28.9$\pm$1.3 & 16.73$\pm$0.05 & 16.97$\pm$0.06 & 16.987$\pm$0.006 & -5.5$\pm$0.2 \\
664.8320040 & 2024 Dec 20 & 5 / 0.007 & 208 / 0.603 & 2395.9 & 139 & 4.7 & 0.2$\pm$9.7 & 9.5$\pm$1.3 & 16.82$\pm$0.05 & 17.02$\pm$0.04 & 17.021$\pm$0.006 & -6.3$\pm$0.2 \\
665.8295021 & 2024 Dec 21 & 5 / 0.095 & 208 / 0.716 & 1978.0 & 134 & 4.7 & 26.4$\pm$9.7 & -8.6$\pm$1.4 & 16.90$\pm$0.05 & 17.13$\pm$0.05 & 17.072$\pm$0.006 & -6.3$\pm$0.2 \\
666.7968469 & 2024 Dec 22 & 5 / 0.180 & 208 / 0.826 & 2111.8 & 141 & 4.4 & 58.7$\pm$9.2 & -29.8$\pm$1.3 & 17.00$\pm$0.05 & 17.12$\pm$0.04 & 17.064$\pm$0.006 & -6.2$\pm$0.2 \\
667.9460719 & 2024 Dec 23 & 5 / 0.282 & 208 / 0.956 & 2384.8 & 138 & 4.6 & 30.9$\pm$9.8 & -55.9$\pm$1.4 & 16.82$\pm$0.06 & 17.03$\pm$0.06 & 17.034$\pm$0.006 & -5.9$\pm$0.2 \\
\hline
712.7557814 & 2025 Feb 06 & 9 / 0.230 & 214 / 0.028 & 2407.1 & 144 & 4.4 & 33.8$\pm$9.2 & -42.7$\pm$1.4 & 16.91$\pm$0.06 & 17.08$\pm$0.06 & 17.043$\pm$0.006 & -6.0$\pm$0.2 \\
713.7517826 & 2025 Feb 07 & 9 / 0.317 & 214 / 0.140 & 2407.1 & 140 & 4.5 & 30.9$\pm$10.1 & -54.5$\pm$1.4 & 16.51$\pm$0.07 & 17.01$\pm$0.05 & 17.018$\pm$0.006 & -5.7$\pm$0.2 \\
714.7552796 & 2025 Feb 08 & 9 / 0.406 & 214 / 0.254 & 2407.1 & 147 & 4.3 & 46.0$\pm$9.2 & -36.6$\pm$1.4 & 16.42$\pm$0.06 & 16.94$\pm$0.05 & 16.912$\pm$0.006 & -5.3$\pm$0.2 \\
716.7313956 & 2025 Feb 10 & 9 / 0.580 & 214 / 0.478 & 2407.1 & 150 & 4.2 & 55.1$\pm$8.8 & -3.4$\pm$1.4 & 16.58$\pm$0.05 & 17.05$\pm$0.04 & 17.005$\pm$0.006 & -6.1$\pm$0.2 \\
717.7468824 & 2025 Feb 11 & 9 / 0.669 & 214 / 0.593 & 2407.1 & 143 & 4.4 & 82.3$\pm$9.7 & 5.4$\pm$1.4 & 16.52$\pm$0.08 & 17.01$\pm$0.06 & 16.940$\pm$0.006 & -6.9$\pm$0.2 \\
718.7253109 & 2025 Feb 12 & 9 / 0.756 & 214 / 0.703 & 2407.1 & 147 & 4.2 & 100.9$\pm$8.8 & 18.8$\pm$1.4 & 16.54$\pm$0.05 & 16.99$\pm$0.05 & 16.946$\pm$0.006 & -6.7$\pm$0.2 \\
719.7597841 & 2025 Feb 13 & 9 / 0.847 & 214 / 0.821 & 2407.1 & 140 & 4.5 & 62.9$\pm$9.9 & 23.4$\pm$1.4 & 16.61$\pm$0.08 & 17.02$\pm$0.05 & 16.971$\pm$0.006 & -6.1$\pm$0.2 \\
720.7188669 & 2025 Feb 14 & 9 / 0.931 & 214 / 0.929 & 2407.1 & 99 & 6.5 & -17.6$\pm$15.2 & 15.7$\pm$2.4 & 16.87$\pm$0.08 & 17.07$\pm$0.08 & 17.034$\pm$0.010 & -5.8$\pm$0.2 \\
721.8351858 & 2025 Feb 15 & 10 / 0.030 & 215 / 0.055 & 2407.1 & 128 & 5.0 & -47.0$\pm$10.2 & -2.4$\pm$1.5 & 16.89$\pm$0.05 & 17.12$\pm$0.06 & 17.038$\pm$0.007 & -5.5$\pm$0.2 \\
724.7468078 & 2025 Feb 18 & 10 / 0.286 & 215 / 0.385 & 2407.1 & 145 & 4.2 & 14.7$\pm$9.2 & -58.7$\pm$1.5 & 16.78$\pm$0.05 & 17.06$\pm$0.06 & 16.998$\pm$0.006 & -5.3$\pm$0.2 \\
725.8192773 & 2025 Feb 19 & 10 / 0.381 & 215 / 0.506 & 2407.1 & 135 & 4.7 & 53.9$\pm$10.4 & -45.3$\pm$1.6 & 16.62$\pm$0.06 & 16.98$\pm$0.08 & 16.951$\pm$0.007 & -5.7$\pm$0.2 \\
\hline
\end{tabular}}
\tablefoot{For each visit, we list the barycentric Julian date BJD, the UT date, the rotation cycle c and phase $\phi$ (assuming $\Prot=11.35$~d, see Sec.~\ref{sec:zdi}, and counting from BJD 2460608, i.e., prior to our first SPIRou observation), the orbital cycle o and phase $\psi$ \citep[using the ephemeris of][i.e., \Porb~=~8.834976~d and BJD0~=~2458821.8257]{Barber2024}, the total observing time t$_{\rm exp}$, the peak SNR in the spectrum (in the $H$ band) per 2.3~\kms\ pixel, the noise level in the LSD Stokes $V$ profile, the estimated \Bl\ with error bars, the differential temperature $dT$ with error bars, the nightly averaged RVs and corresponding error bars derived from Stokes $I$ LSD profiles of atomic and CO lines, the RVs and corresponding error bars from the LBL analysis, and the measured Pa-$\beta$ EWs. }
\label{tab:log}
\end{table*}

\section{Gaussian Process regression}
\label{sec:appB}

To investigate rotational modulation of our \Bl\ and $dT$ data, allowing for temporal evolution of the modulation pattern, we use quasi-periodic GPR.  This is achieved by finding out the hyper parameters of the covariance function that best describes our data, arranged in a vector denoted $\bf y$.  The quasi-periodic covariance function $c(t,t')$ we use in this purpose is as follows: 
\begin{eqnarray}
c(t,t') = \theta_1^2 \exp \left( -\frac{(t-t')^2}{2 \theta_3^2} -\frac{\sin^2 \left( \frac{\pi (t-t')}{\theta_2} \right)}{2 \theta_4^2} \right) 
\label{eq:covar}
\end{eqnarray}
where $\theta_1$ is the amplitude of the GP, $\theta_2$ its recurrence period (directly linked to \Prot), $\theta_3$ the
evolution timescale (in d) on which the shape of the modulation changes, and $\theta_4$ a smoothing parameter describing the amount of harmonic complexity needed to describe the data \citep{Haywood2014,Rajpaul2015}.  
We then select the set of hyper parameters that yields the highest likelihood $\mathcal{L}$, defined by:
\begin{eqnarray}
2 \log \mathcal{L} = -n \log(2\pi) - \log|{\bf C+\Sigma+S}| - {\bf y^T} ({\bf C+\Sigma+S})^{-1} {\bf y}
\label{eq:llik}
\end{eqnarray}
where $\bf C$ is the covariance matrix for our epochs, $\bf \Sigma$ the diagonal variance matrix associated with $\bf y$, and
${\bf S}=\theta_5^2 {\bf J}$ ($\bf J$ being the identity matrix) the contribution from an additional white noise source used as a fifth
hyper-parameter $\theta_5$.  We finally use a MCMC process to explore the hyper-parameter domain, yielding posterior distributions and error bars for each of them.  The marginal logarithmic likelihood $\log \mathcal{L}_M$ of a given solution is computed using the approach of \citet{Chib2001} as in, e.g., \citet[][]{Haywood2014}.

\section{Magnetic field and temperature fluctuations} 
\label{sec:appZ}

\begin{table}
\caption[]{Results of our GPR MCMC modeling of the \Bl\ (top rows) and $dT$ (bottom rows) curves of \irasstar. }
\centering                      
\resizebox{\linewidth}{!}{
\begin{tabular}{ccc}
\hline
Parameter   & value & Prior   \\
\hline
\Bl &&  \\
Rec.\ period (d)     & $11.35\pm0.04$   & Gaussian (11.35, 2.0) \\
GP amplitude (G)     & $48\pm17$        & mod Jeffreys ($\sigma_{\Bl}$) \\
Evol.\ timescale (d) & $97\pm30$        & log Gaussian ($\log$ 100, $\log$ 2) \\
Smoothing            & $0.43\pm0.09$    & Uniform  (0, 3) \\
White noise (G)      & $3.5\pm2.3$      & mod Jeffreys ($\sigma_{\Bl}$) \\
Rms (G)              & 6.1              & \\
$\chisqr$            & 0.36             &  \\
\hline
$dT$ &&  \\
Rec.\ period (d)     & $11.28\pm0.05$   & Gaussian (11.35, 2.0) \\
GP amplitude (K)     & $30\pm10$        & mod Jeffreys ($\sigma_{dT}$) \\
Evol.\ timescale (d) & $97\pm32$        & log Gaussian ($\log$ 100, $\log$ 2) \\
Smoothing            & $0.53\pm0.12$    & Uniform  (0, 3) \\
White noise (K)      & $1.5\pm0.8$      & mod Jeffreys ($\sigma_{dT}$) \\
Rms (K)              & 1.6              & \\
$\chisqr$            & 1.2              & \\
\hline
\end{tabular}}
\tablefoot{For each hyper parameter, we list the fitted value along with the corresponding error bar, as well as the assumed prior.  The knee of the modified Jeffreys prior is set to the median error bars of our \Bl\ and $dT$ estimates (i.e., 10~G, 40~G and 1.4~K respectively).}
\label{tab:gpr}
\end{table}

\begin{figure*}[h!]
    \centering
    \includegraphics[width=0.8\textwidth]{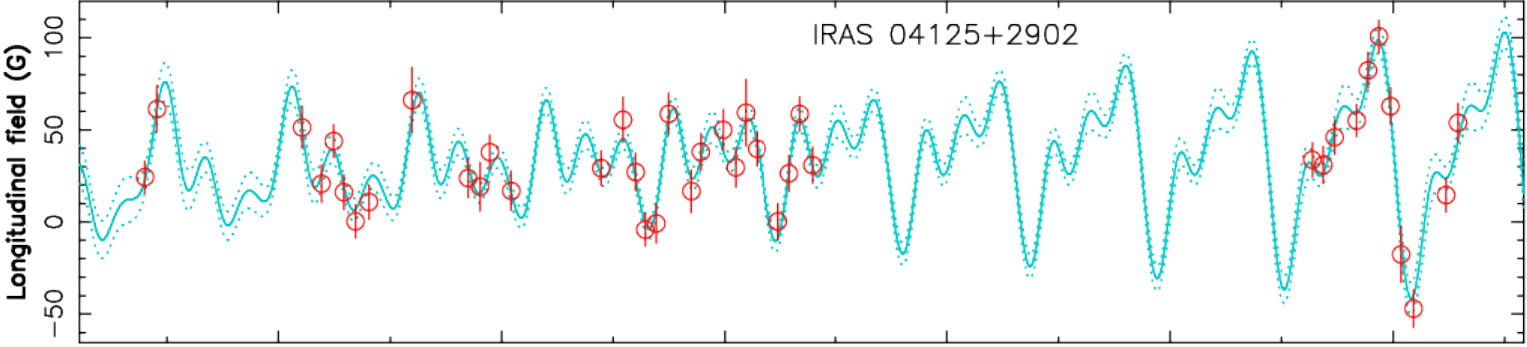}\vspace{-0.9mm}
    \includegraphics[width=0.8\textwidth]{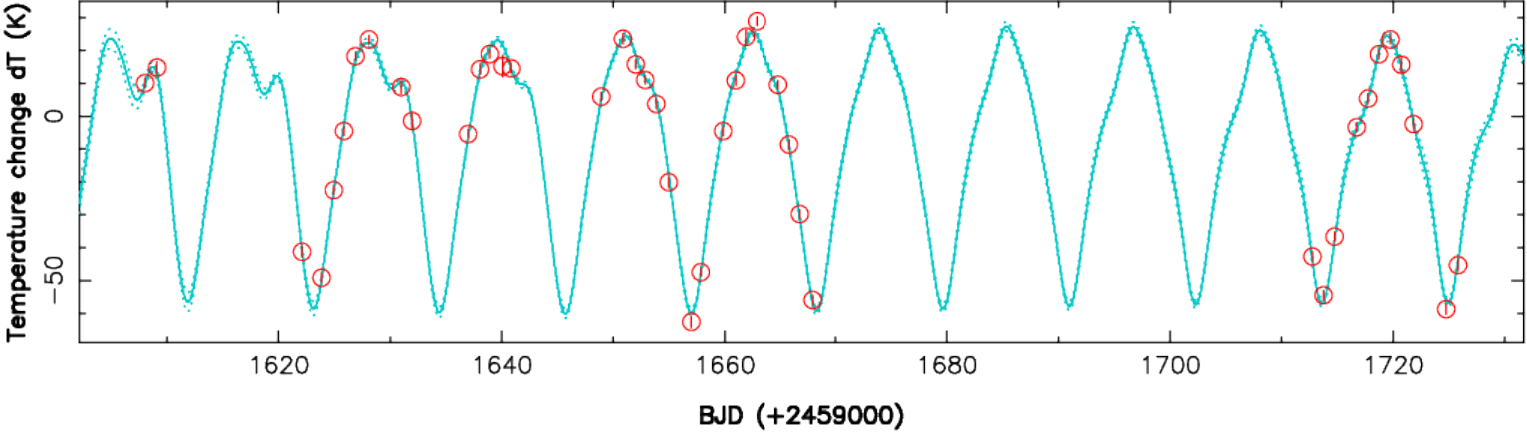}
    \caption{Longitudinal magnetic field \Bl\ (top panel) and temperature changes $dT$ (bottom panel) of \irasstar\ (red open circles) as measured with SPIRou, and quasi-periodic GPR fit to the data (cyan full line) with corresponding 68\% confidence intervals (cyan dotted lines). }
    \label{fig:bell}
\end{figure*}

\section{Zeeman Doppler imaging}
\label{sec:appC}

ZDI allows one to reconstruct the brightness distribution and the large-scale magnetic field at the surface of a rotating star from phase-resolved sets of Stokes $I$ and $V$ LSD profiles and contemporaneous photometry.  This is performed through an iterative process, starting from a small magnetic field and a featureless brightness map and progressively adding information at the surface of the star until the modeled profiles and photometry match the observed ones at the required level, usually a unit reduced chi-square $\chisqr\simeq1$.  As this inversion problem is ill-posed, regularization is required to ensure a unique solution;  in our case, we use the principles of maximum entropy image reconstruction \citep{Skilling1984} to select the image that features minimal information among those matching the data.

In practice, we model the stellar surface as a grid of 5000 cells, whose spectral contributions are computed using Unno-Rachkovsky's analytical solution of the polarized radiative transfer equation in a plane-parallel Milne Eddington atmosphere \citep{Landi2004}, with a local profile centered on 1750~nm and featuring a Doppler width and Land\'e factor of 2.5~\kms\ and 1.2 respectively (as in previous studies).  We then compute the synthetic profiles at the observed rotation cycles by summing the contributions of all grid cells, taking into account their respective velocities with respect to the observer.  Relative brightness at the surface of the star is simply described as a series of independent values, whereas the large-scale magnetic field is expressed as a spherical harmonic expansion in which the poloidal and toroidal components of the vector field are parametrized with three independent sets of complex coefficients \citep{Donati2006, Finociety2022}.

\begin{figure}[ht!]
    \centering
    \includegraphics[width=0.48\linewidth]{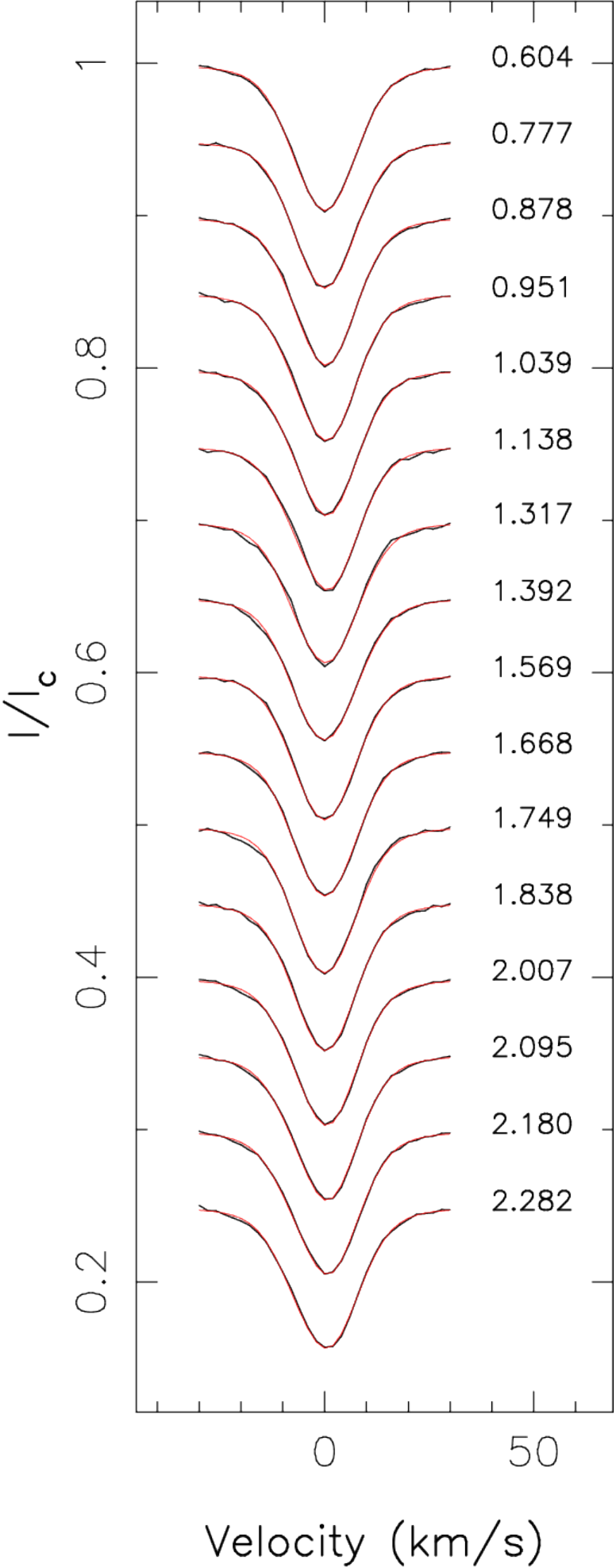}\hspace{2mm}\includegraphics[width=0.48\linewidth]{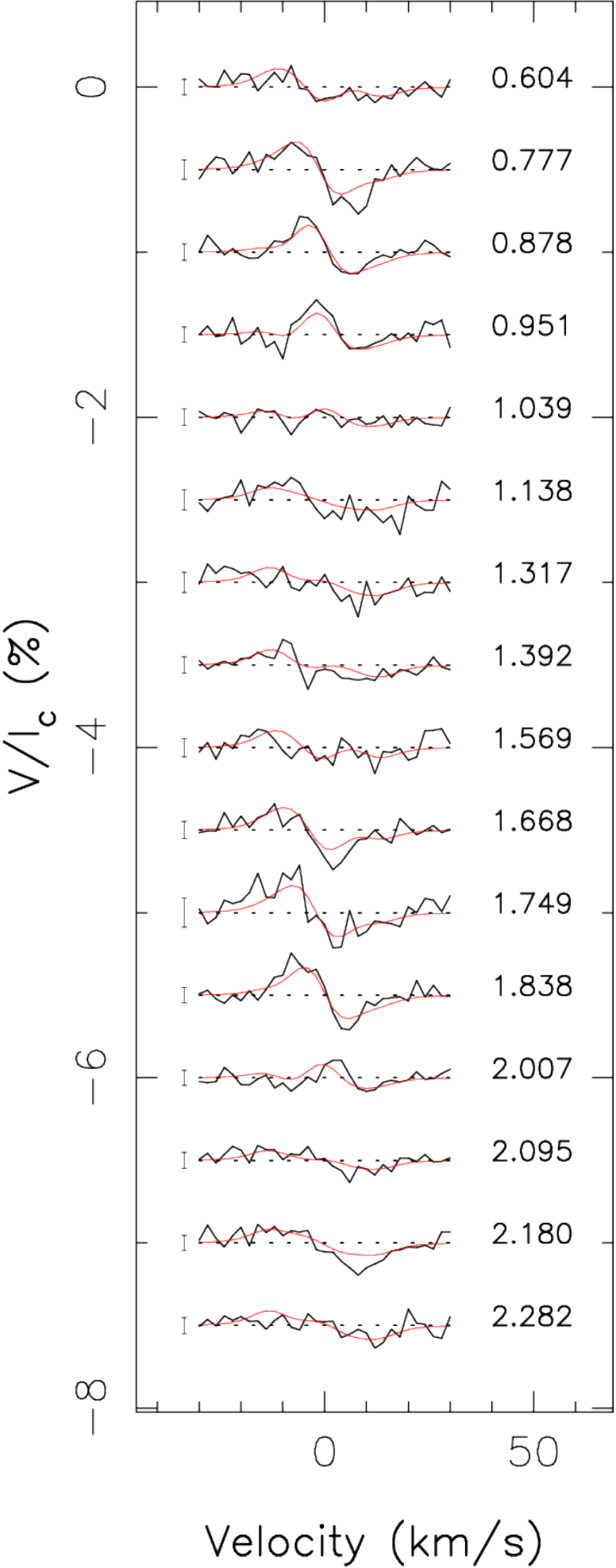}\vspace{2mm}
    \includegraphics[width=0.95\linewidth]{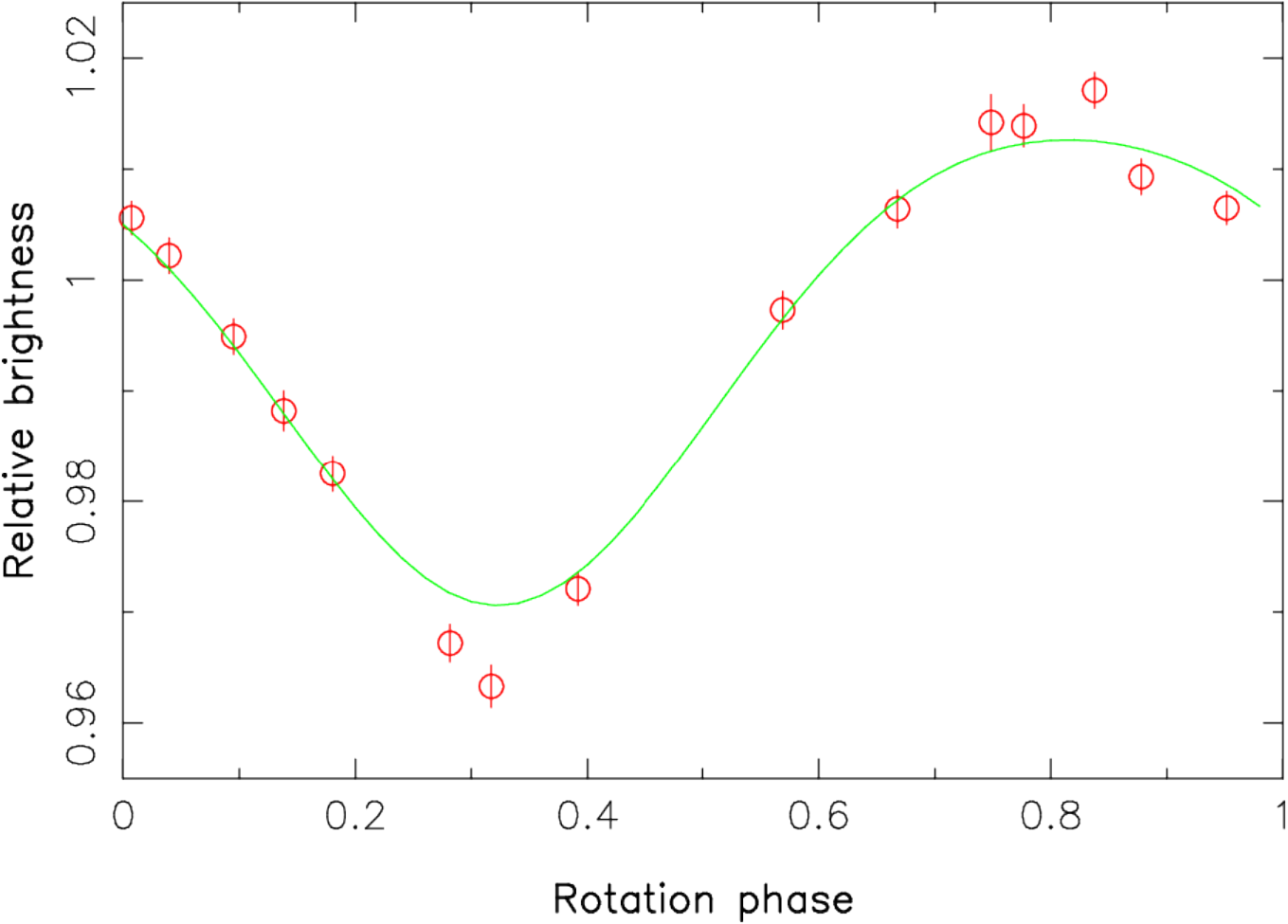}
    \caption{Top panels: Observed (black line) and ZDI modeled (red line) LSD Stokes $I$ (left) and $V$ (right) profiles of \irasstar\ for our 2024 December data set.  {\emr Both fits to the LSD Stokes $I$ and $V$ profiles yield $\chisqr=1$, i.e., are consistent with the data down to photon noise}.  Rotation cycles (counting from 3, see Table~\ref{tab:log}) are indicated to the right of the LSD profiles, while $\pm$1$\sigma$ error bars are added to the left of the Stokes $V$ signatures.    Bottom panel: photometry derived from $dT$ (red circles) and ZDI fit (green line), {\emr reproducing most of the observed variability caused by large surface brightness features}. }
    \label{fig:fitIV}
\end{figure}

\begin{figure*}[ht!]
    \centering
    \includegraphics[width=0.85\textwidth]{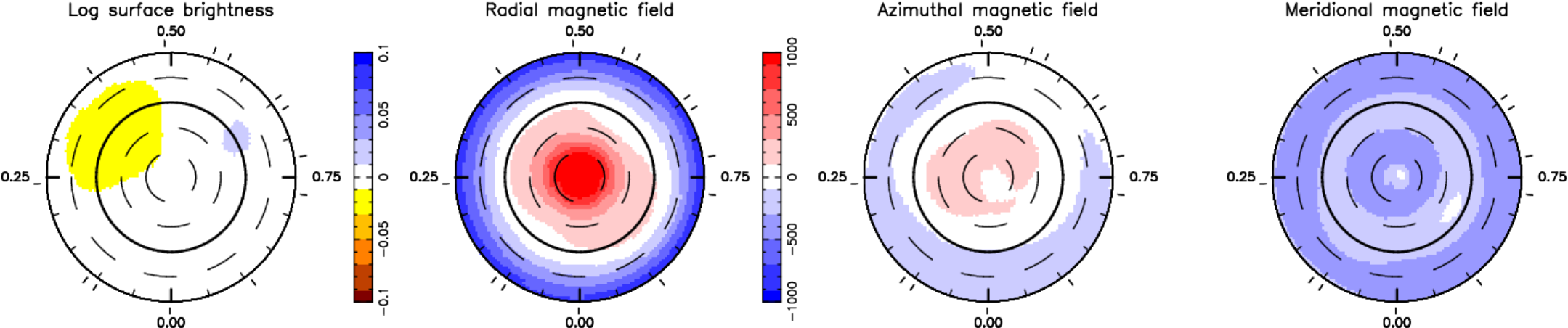}\vspace{2mm}
    \includegraphics[width=0.85\textwidth]{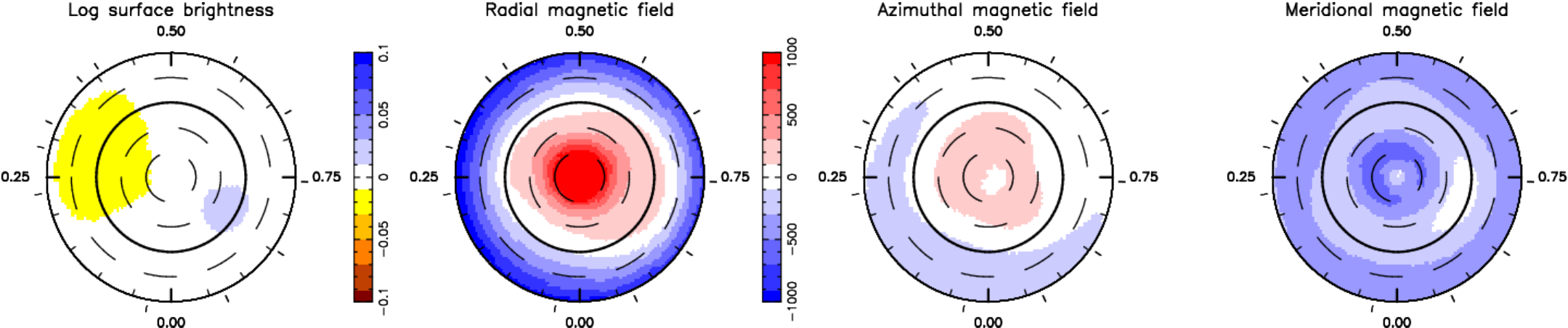}

    \caption{{\emr Same as Fig.~\ref{fig:mapB} for the other two subsets (October+November 2024 and February 2025 in the top and bottom row respectively), showing the limited temporal evolution in the brightness and magnetic maps over the 118~d of our monitoring.}}  
    \label{fig:mapB2}
\end{figure*}

\section{GPR fit and periodograms of RV curves}
\label{sec:appD}

\begin{table}
\caption[]{Same as Table~\ref{tab:gpr} for our joint modeling of the RV curves of atomic and CO lines of \irasstar.  }
\centering                      
\resizebox{\linewidth}{!}{
\begin{tabular}{ccc}
\hline
Parameter   & value & Prior   \\
\hline
Rec.\ period (d)     & $11.32\pm0.05$   & Gaussian (11.35, 2.0) \\
K (\ms)              & $K=19^{+13}_{-8}$ & mod Jeffreys ($\sigma_{RV}$) \\
\hline
atomic lines &&  \\
GP amplitude (\ms)   & $174\pm38$       & mod Jeffreys ($\sigma_{RV}$) \\
Evol.\ timescale (d) & 100              & fixed \\
Smoothing            & 0.4              & fixed \\
White noise (\ms)    & $32\pm20$        & mod Jeffreys ($\sigma_{RV}$) \\
\hline
CO lines &&  \\
GP amplitude (\ms)   & $69\pm23$        & mod Jeffreys ($\sigma_{RV}$) \\
Evol.\ timescale (d) & 140              & fixed \\
Smoothing            & 0.5              & fixed \\
White noise (\ms)    & $19\pm19$        & mod Jeffreys ($\sigma_{RV}$) \\
\hline
Rms (\ms)            & 45               & \\
$\chisqr$            & 0.67             & \\
\hline
\end{tabular}}
\tablefoot{The evolution timescales and smoothing factors, weakly constrained by the data, were set to an optimum derived from a preliminary run, with no impact on the result.  The knee of the modified Jeffreys prior is set to the median RV error bars (54~\ms\ for both line sets).}
\label{tab:gprv}
\end{table}

\begin{table}
\caption[]{Same as Table~\ref{tab:gprv} for the LBL RVs.}
\centering                      
\resizebox{\linewidth}{!}{
\begin{tabular}{ccc}
\hline
LBL &&  \\
Rec.\ period (d)     & $11.46\pm0.25$   & Gaussian (11.35, 2.0) \\
K (\ms)              & $8.6^{+9.6}_{-4.5}$ & mod Jeffreys ($\sigma_{RV}$) \\
GP amplitude (\ms)   & $76\pm15$       & mod Jeffreys ($\sigma_{RV}$) \\
Evol.\ timescale (d) & $18\pm4$         & log Gaussian ($\log$ 20, $\log$ 2) \\
Smoothing            & $0.36\pm0.07$    & Uniform  (0, 3) \\
White noise (\ms)    & $12\pm9$        & mod Jeffreys ($\sigma_{RV}$) \\
Rms (\ms)            & 4               & \\
$\chisqr$            & 0.30            & \\
\hline
\end{tabular}}
\tablefoot{All GPR parameters are free to vary and the knee of the modified Jeffreys prior is set 6~\ms.}
\label{tab:gprv2}
\end{table}

\begin{figure*}[ht!]
    \centering
    \includegraphics[width=0.8\textwidth]{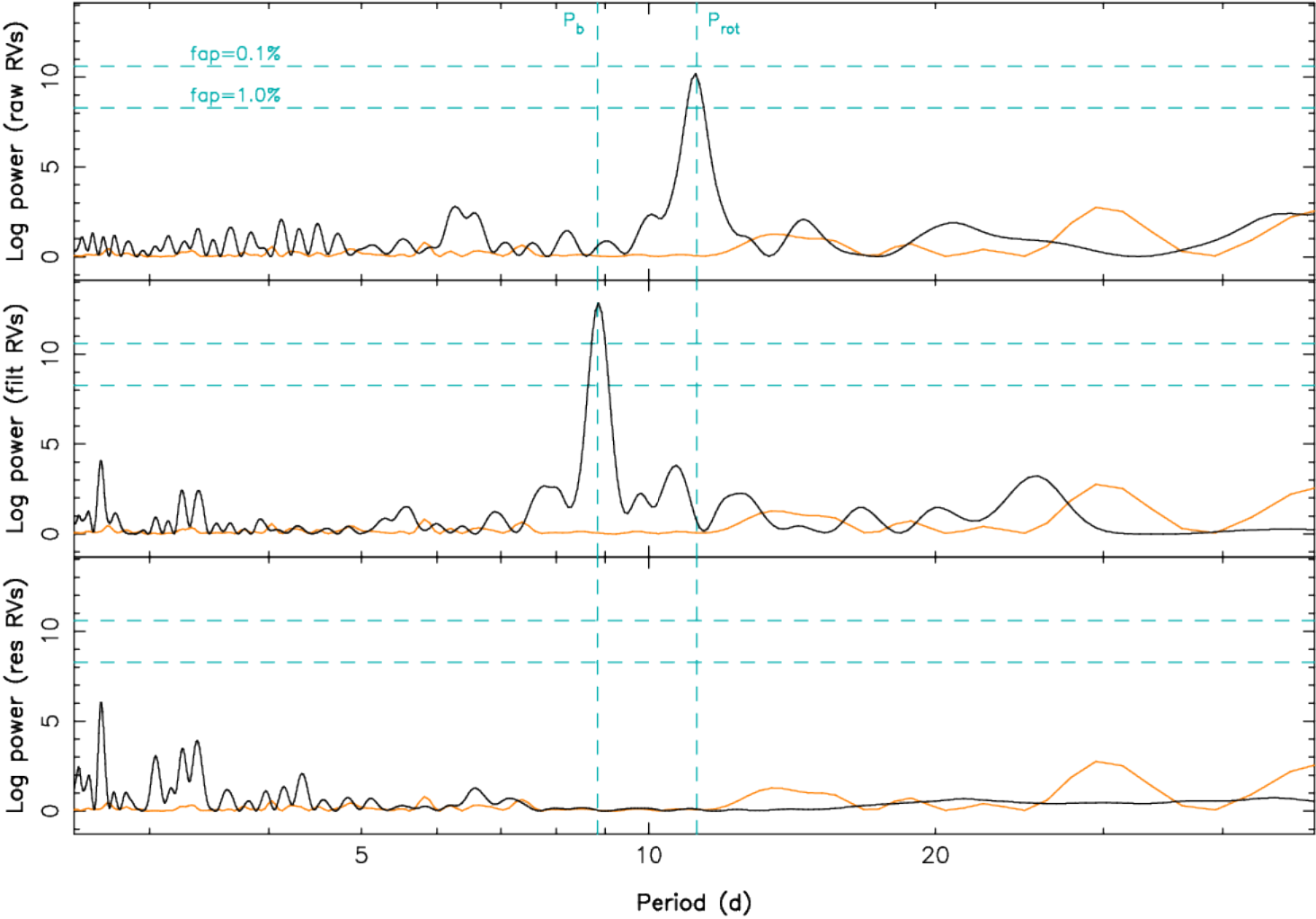}
    \caption{Periodograms of the raw (top plot), filtered (middle plot) and residual (bottom plot) LBL RVs of \irasstar. The cyan vertical dashed lines trace the derived \Prot\ and known orbital period $P_b$ of \irasstar~b, whereas the horizontal dashed lines indicate the 1 per cent and 0.1 per cent false alarm probabilities in the periodograms of our RV data. The orange curve depicts the periodogram of the window function.  }
    \label{fig:per}
\end{figure*}

\begin{figure*}[ht!]
    \centering
    \includegraphics[width=0.48\textwidth]{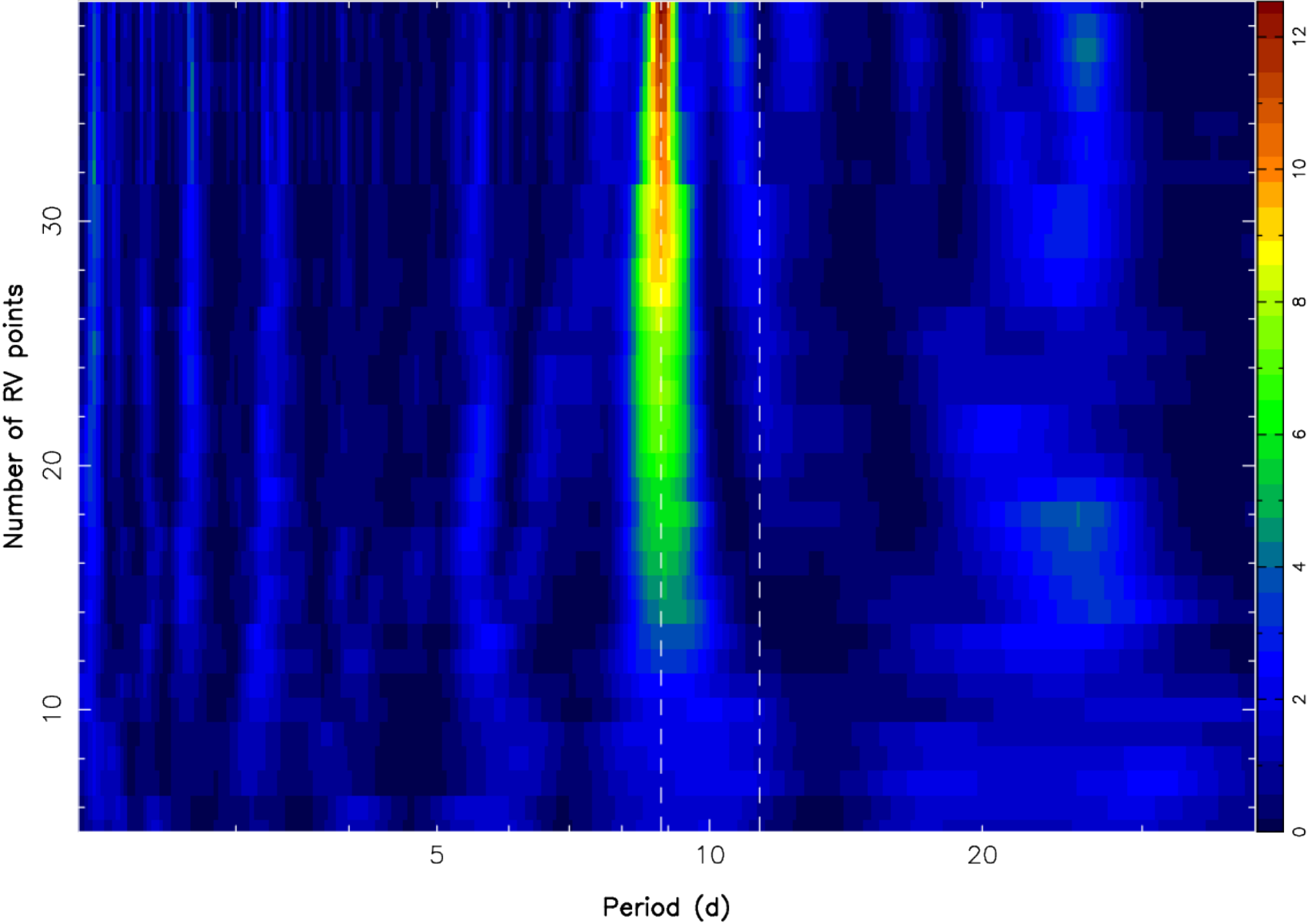}\hspace{4mm}
    \includegraphics[width=0.48\textwidth]{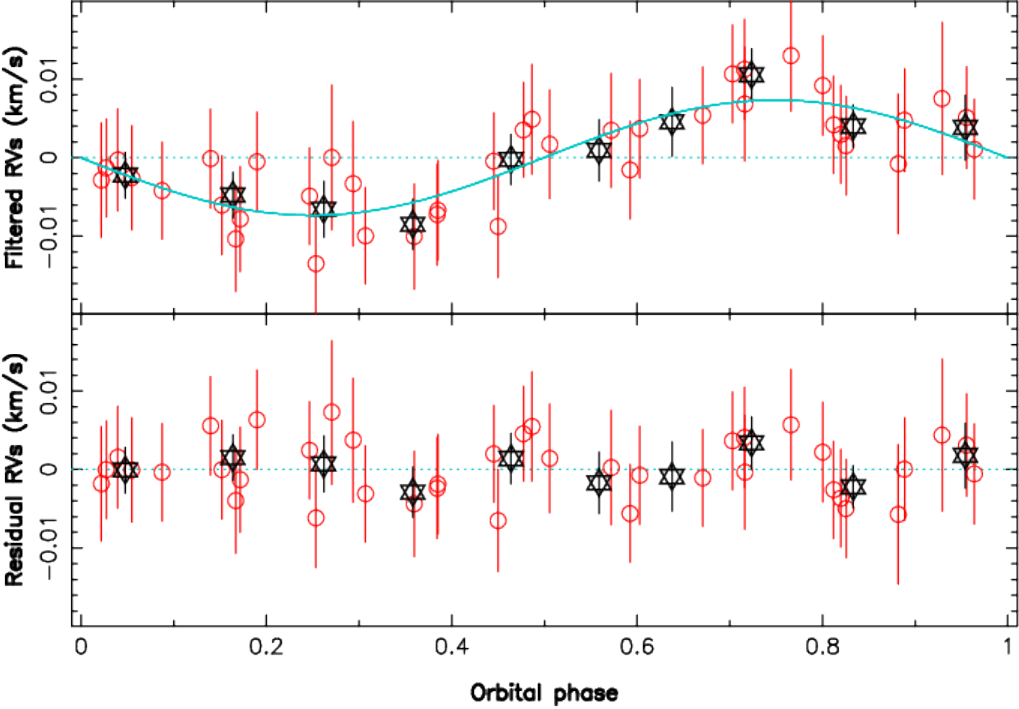}
    \caption{Stacked periodograms (left panel) and phase-folded (right panel, top plot) filtered RVs of \irasstar.  The stacked periodogram (with colors coding periodogram power) shows a marginal signal at the orbital period that strengthens as RV points are added to the analysis.  In the right panel{\emr , the black stars show the RVs averaged over phase bins of 10\%, with the phase-folded residuals displayed in the bottom plot}.  }
    \label{fig:per2}
\end{figure*}

\section{Paschen~$\beta$ and Brackett~$\gamma$ lines}

\begin{figure*}[ht!]
    \centering
    \includegraphics[width=0.495\textwidth]{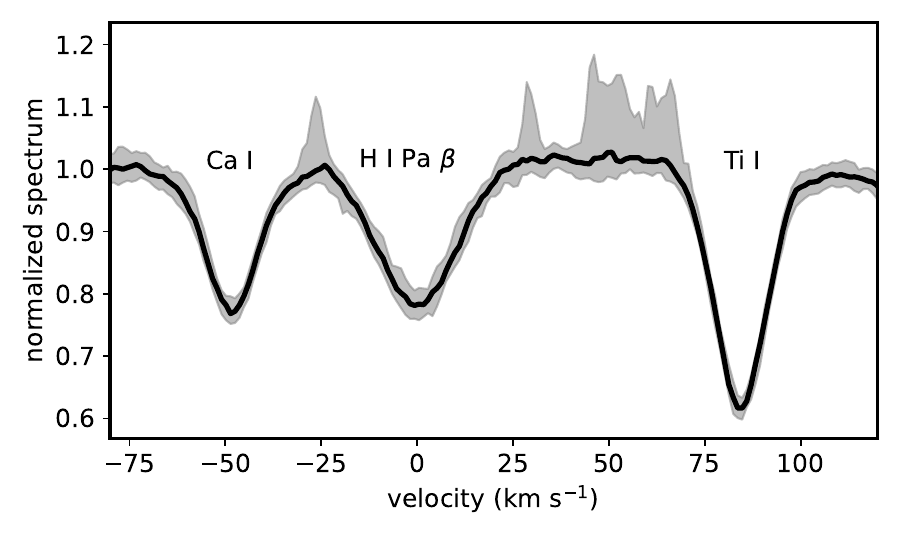}\hspace{0mm}
    \includegraphics[width=0.495\textwidth]{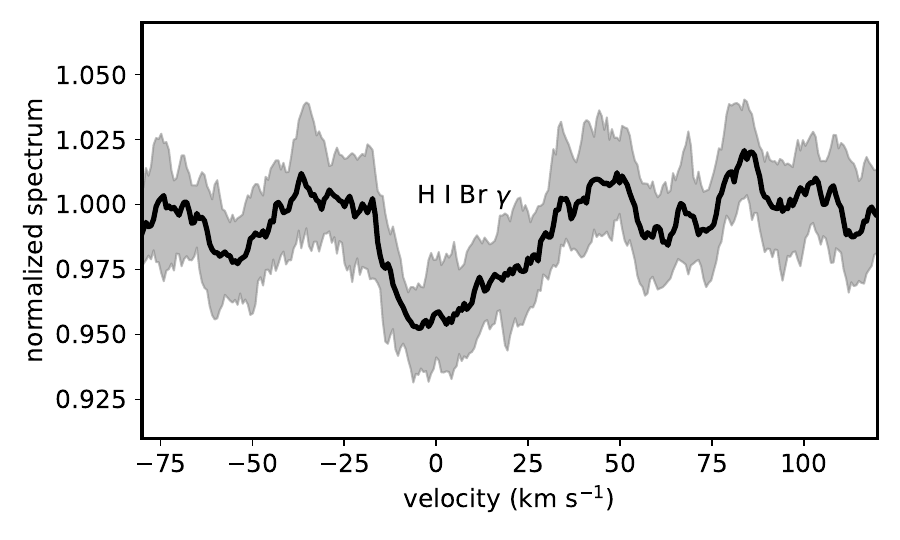}
    \caption{SPIRou median and rms spectrum of \irasstar\ in the vicinity of the Paschen~$\beta$ (left panel) and Brackett~$\gamma$ (right panel) of H\,I.    Marginally larger variability shows up in Paschen~$\beta$ relative to the neighboring lines of Ca\,I and Ti\,I.  Residuals of OH emission lines are responsible for the positive features in the 68\% scatter. }
    \label{fig:pabeta}
\end{figure*}

\end{appendix}
\end{document}